\documentclass[a4paper,11pt]{article}
\usepackage{amsmath}
\usepackage{amssymb}
\usepackage{fullpage}
\usepackage{verbatim}

\usepackage[T1,OT1]{fontenc} 

\usepackage{graphicx}

\allowdisplaybreaks[1]

\usepackage{hyperref}
\usepackage{cite}

\renewcommand{\d}{\mathrm{d}}
\newcommand{\e}{\mathrm{e}}
\newcommand{\w}{\wedge}
\newcommand{\ti}{\tilde}

\newcommand{\ext}{\text{ext}}
\newcommand{\inte}{\text{int}}
\newcommand{\f}[2]{\frac{#1}{#2}}
\newcommand{\diff}[2]{\f{\d #1}{\d #2}}
\newcommand{\de}[2]{\f{\delta #1}{\delta #2}}
\newcommand{\Z}{{\mathbb Z}}

\begin{document}

\numberwithin{equation}{section}

\thispagestyle{empty}

\begin{flushright}
\small ITP-UH-02/13\\
\normalsize
\end{flushright}
\vspace{1cm}

\begin{center}

{\LARGE \bf Cosmological Constant, Near Brane Behavior} 

\vskip4mm
{\LARGE \bf and Singularities}

\vskip3mm
 {\LARGE \bf }

\vspace{1cm}
{\large Fri\fontencoding{T1}\selectfont \dh \fontencoding{OT1}\selectfont rik Freyr Gautason${}^1$, Daniel Junghans${}^{1,2}$ and Marco Zagermann${}^{1}$}\\

\vspace{1cm}
\vspace{.15 cm}  ${}^1$ {Institut f{\"u}r Theoretische Physik \&\\
Center for Quantum Engineering and Spacetime Research\\
Leibniz Universit{\"a}t Hannover, Appelstra{\ss}e 2, 30167 Hannover, Germany}\\

\vspace{0.5cm}
\vspace{.15 cm}  ${}^2$ Department of Physics \& Institute for Advanced Study,\\
Hong Kong University of Science and Technology, Hong Kong 

\vspace{1cm}
{\upshape\ttfamily fridrik.gautason, daniel.junghans, marco.zagermann@itp.uni-hannover.de}\\

\vspace{1cm}
\begin{abstract}
\noindent 
We show that the classical cosmological constant in type II flux compactifications
can be written as a sum of 
terms from the action of localized sources plus a specific contribution from
 non-trivial background fluxes.
 Exploiting two global scaling symmetries of the classical supergravity action, we find that the flux contribution can in many interesting cases be set to zero such that the cosmological constant is fully determined by the boundary conditions of the fields in the near-source region. This generalizes and makes more explicit previous arguments in the literature. We then discuss the problem of putting $\overline{\textrm{D}3}$-branes at the tip of the Klebanov-Strassler throat glued to a compact space in type IIB string theory so as to engineer a de Sitter solution.
Our result for the cosmological constant and a simple global argument indicate that 
inserting a fully localized and backreacting $\overline{\textrm{D}3}$-brane into such a background yields a singular energy density
 for the NSNS and RR 3-form field strengths at the $\overline{\textrm{D}3}$-brane. This argument does not rely on partial smearing of the $\overline{\textrm{D}3}$-brane or a linearization of field equations, but on a few general assumptions that we also discuss carefully. 
\end{abstract}

\end{center}

\newpage
\tableofcontents
\vspace{0.5cm}

\section{Introduction}

A better understanding of string compactifications involving localized sources such as D-branes or orientifold planes is an important task for string phenomenology. 
Unfortunately, for most scenarios a full solution to the ten-dimensional equations of motion seems to be out of reach, even in the supergravity approximation, 
because the involved differential equations are  too complex. On the other hand, commonly used procedures for simplifying this task, such as a smearing of the localized 
sources over the compact space, may introduce their own problems and need not necessarily capture essential features of the true solution 
(see e.g. \cite{Douglas:2010rt,Blaback:2010sj,Blaback:2011nz}). 
It would therefore be desirable to be able to compute important observables such as the cosmological constant without having to know the full ten-dimensional dynamics 
or rely on simplifications such as smearing.

In the first part of this work, we will show, in the context of type II supergravity coupled to D-branes and O-planes, that such a method often exists, building upon previous 
work that had already pointed towards this possibility \cite{Aghababaie:2003ar,Burgess:2011rv,Gautason:2012tb}. In particular, we will 
argue that the cosmological constant, $\Lambda$,  can often be expressed as a sum of terms that are due to the action of localized sources,
\begin{equation}
\Lambda \propto \sum_{p} c_p \left({S^{(p)}_\textrm{DBI} + S^{(p)}_\textrm{CS}}\right),
\end{equation}
where $S^{(p)}_\textrm{DBI}$ and $S^{(p)}_\textrm{CS}$ are the on-shell evaluated DBI and Chern-Simons actions of the D$p$-branes and/or O$p$-planes present in the corresponding supergravity solution, and $c_p$ are $p$-dependent constants. Thus, in compactification scenarios where our reasoning holds, $\Lambda$ is entirely specified by the classical boundary conditions of some of the bulk fields at the positions of the sources and independent of the details of the ten-dimensional bulk dynamics.

Such a property was noticed before in \cite{Aghababaie:2003ar,Burgess:2011rv}.
Using a single scaling symmetry of the action of different supergravity theories, the authors were able to relate $\Lambda$ to 
boundary terms involving the supergravity fields that have to be evaluated in the near-source region. They pointed out, however, that topologically nontrivial background fluxes can also give contributions that arise from the patching of gauge charts. The explicit evaluation of these subtle flux and all source contributions together with their gauge dependence for a general type II compactification make up the first part of our paper. Furthermore, we show that the flux contributions actually vanish in many interesting examples such that the only contributions to the cosmological constant are due to the action of D-branes and O-planes.

From a somewhat different angle, also the results of \cite{Gautason:2012tb} suggested such a behavior. 
There, it was shown that the cosmological constant in solutions of perturbative heterotic string theory is zero to all orders in $\alpha^\prime$, 
unless one introduces spacetime-filling fluxes or considers string loop or non-perturbative corrections. Since the argument only used the scaling properties
of the effective potential with respect to the dilaton, it was then conjectured that a similar reasoning should also be applicable for the type II string,
with the exception that then also D-brane and O-plane sources should give a contribution to $\Lambda$. For classical solutions of type II supergravity, this suggests
that, in absence of spacetime-filling flux, any non-zero contribution to $\Lambda$ must be generated by terms that are due to localized sources. It turns out, however, that the intuitive scaling argument of \cite{Gautason:2012tb} is complicated in the type II string by a subtlety related to the RR 
fields: in a frame, where the bulk action scales uniformly with the dilaton, non-trivial couplings of the RR potentials with derivatives of the dilaton of 
the form $\d \phi \w C \w F$ arise. These couplings are only present in the type II string, not in the heterotic string. In the presence of background fluxes,
they can be shown to yield non-zero contributions to $\Lambda$, thus spoiling the argument sketched above.

As we will show in this paper, however, it remains true in many cases that $\Lambda$ is completely determined by a sum of source terms. The reason is that classical type II (and also heterotic)
supergravity exhibits a two-parameter scaling symmetry, related to the dilaton scaling and the mass scaling of
 the classical action \cite{witten:1985xb,Burgess:1985zz}. Both the scaling 
symmetry
exploited in \cite{Burgess:2011rv} and the one implicitly used in \cite{Gautason:2012tb} are special cases of this more general symmetry. As we will show below,
it ensures that one can often find a particular combination of the equations of motion such that all bulk terms are eliminated from the equation determining $\Lambda$,
leaving a contribution entirely from localized sources. The cosmological constant is then indeed given by a sum of source terms as initially claimed. More precisely, this can be shown
to hold for maximally symmetric compactifications of type II supergravity involving sources of arbitrary dimension and at most $H$ flux and one type of RR flux.

In the second part of this work, we discuss an application of our result to the idea of placing $\overline{\textrm{D}3}$-branes at the bottom of the Klebanov-Strassler solution 
\cite{Klebanov:2000hb,Giddings:2001yu,Kachru:2002gs}, a setup that has been suggested for the construction of meta-stable de Sitter vacua in string theory starting with \cite{Kachru:2003aw}. The backreaction of $\overline{\textrm{D}3}$-branes on the Klebanov-Strassler geometry
has recently been subject of intense discussions \cite{DeWolfe:2008zy,McGuirk:2009xx,Bena:2009xk,Bena:2011hz,Bena:2011wh,Dymarsky:2011pm,Massai:2012jn,Bena:2012bk,Bena:2012vz,Bena:2012ek,Blaback:2012nf}. Part of this debate 
concerns the computational evidence for a singularity in fields that do not directly couple to the anti-branes as it emerged in several approaches.

More precisely, the presence of this singularity has so far been demonstrated in simplified setups that use certain approximations. In earlier works on the subject, 
this involved a partial smearing of the branes and a linearization of the equations of motion around the Klebanov-Strassler background \cite{McGuirk:2009xx,Bena:2009xk,Bena:2011hz,Bena:2011wh}.
 \cite{Dymarsky:2011pm} therefore also discusses the possibility that the singularity might just be an artifact of perturbation theory and disappear in the full setup (see however \cite{Massai:2012jn}).
Although it could recently be shown in \cite{Bena:2012bk} that also the non-linear equations of motion necessarily lead to a singular solution, the analysis still required partially smeared 
branes. An analysis of the fully localized case could only be carried out for a simplified toy model with $\overline{\textrm{D}6}$-branes \cite{Blaback:2011nz,Blaback:2011pn,Bena:2012tx}, which is related by T-duality to partially smeared $\overline{\textrm{D}3}$-branes on $\mathbb{R}^3 \times T^3$ \cite{Massai:2012jn}. In this simplified setup, it was shown 
that fully localized branes in a non-BPS flux background lead to a singularity in the energy density of the $H$ flux, which is not directly sourced by the $\overline{\textrm{D}6}$-branes. 

As our result from the first part of the paper relates the near-brane behavior of the supergravity fields to the effective cosmological constant, it is natural to try to apply this to $\overline{\textrm{D}3}$-branes in the KS background. We show that under a few assumptions this would indeed be possible and confirm
the presence of a non-standard singularity at the $\overline{\textrm{D}3}$-brane similar to the one discussed before, but now without the approximation of
any smearing and by using the full non-linear supergravity equations.

This paper is organized as follows. In Section \ref{conventions}, we establish our notation and conventions and state the equations of motion of type II supergravity used in the following sections. In Section \ref{cc}, we discuss the two scaling symmetries of classical type II supergravity. We then show that the cosmological constant can be written as a sum of source terms and a term involving topological background fluxes that can in many cases be gauged away by exploiting a combination of the symmetries. In Section \ref{examples}, we present several explicit examples of compactifications of type II supergravity and show how our framework can be applied to them in order to obtain an expression for the cosmological constant in terms of the actions of localized sources. In Section \ref{kklt}, we consider the backreaction of $\overline{\textrm{D}3}$-branes on the Klebanov-Strassler throat 
glued to a compact space in type IIB string theory and discuss under what assumptions our previous results would imply the existence of a singularity in the energy densities of $H$ and $F_3$. We conclude with some comments in Section \ref{concl}.
\\

\section{Type II Supergravity}
\label{conventions}

We start by establishing our notation and conventions.\footnote{We use the conventions of \cite{Koerber:2010bx} except that the sign of $B$ is flipped.} In the tree-level supergravity approximation, the low energy effective action of type II string theory in Einstein frame can be written as
\begin{equation}\label{cc:action}
S = S_\text{bulk} + S_\text{loc}
\end{equation}
with
\begin{equation}\label{cc:bulkaction}
S_\text{bulk} = S_\text{NSNS} + S_\text{RR} = \int\star_{10}\left\{ R - \f{1}{2}|\d\phi|^2 - \f{1}{2}\e^{-\phi}|H|^2 - \f{1}{4}
\sum_n \e^{\tfrac{5-n}{2}\phi}|F_{n}|^2\right\}.
\end{equation}
Here, $R$ is the curvature scalar of the metric $g$, $\star_{10}$ denotes the ten-dimensional Hodge operator associated with $g$,
$\phi$ is the dilaton, $H$ is the NSNS 3-form field strength, and $F_n$ are the RR field strengths.
For an $n$-form $A$, the norm $|A|^2$ is defined by
\begin{equation}
|A|^2\,\,\star_{10} \! 1 =  \frac{1}{n!}A_{\mu_1\ldots\mu_n}A^{\mu_1\ldots\mu_n}\,\,  \star_{10} \! 1 =(\star_{10}A)\wedge A.
\end{equation}
We often consider warped product spaces $\mathcal{M}^{(10)}=\mathcal{M}^{(d)}\times_{w} \mathcal{M}^{(k)}$,
where 
 $\star_d$ and $\star_k$ then denote the Hodge operators of the corresponding warped metric factors. For factorizing forms 
$A_p\wedge B_q$, where 
$A_p$ is a $p$-form on $\mathcal{M}^{(d)}$ and $B_q$ a $q$-form on 
$\mathcal{M}^{(k)}$, these Hodge operators satisfy the useful identity
 $\star_{10}(A_p\wedge B_q)=(-1)^{p(k-q)}(\star_d A_p)\wedge  (\star_k B_q)$.
In general, we have $(\star_{D})^{2} A_p = (-1)^{p(D-p)+t} A_p$ for any $p$-form on a $D$-dimensional manifold with $t$ timelike directions.

Throughout this paper, we use the democratic formulation \cite{Bergshoeff:2001pv}, so that the sum over the RR field strengths in \eqref{cc:bulkaction} also includes the dual fields with $n>5$. The field strengths are related to one another by the duality relations
\begin{equation} \label{rrduality}
\e^{\tfrac{5-n}{2}\phi}F_n  = \star_{10}\, \sigma(F_{10-n}),
\end{equation}
which have to be imposed on-shell. The operator $\sigma$ here acts on an $n$-form $\omega_n$ like
\begin{equation}
\sigma(\omega_n) = (-1)^{\tfrac{n(n-1)}{2}} \omega_n. \label{reversal}
\end{equation}
Also notice that, in \eqref{cc:bulkaction}, we have set $2\kappa_{10}^2=1$, so that the Planck mass has been absorbed into the definition of the metric.

The term $S_\text{loc}$ denotes the action of localized sources corresponding to either D$p$-branes or O$p$-planes and reads\footnote{We do not include the NSNS $2$-form in the DBI action
here, because in all the examples we discuss in detail the sources are either point-like in the internal space or they are wrapped O-planes, so that a $B$-field along the world volume cannot occur. Likewise we do not consider D-branes with world volume fluxes in our examples and hence also omit them in the DBI action. It is easy to check that omitting the NSNS $2$-form in the DBI action does not lead to a missing term in the $H$-equation of motion, because $\delta S_{DBI}/\delta B_{\mu\nu}$ also vanishes if $B$ and $F$ are set to zero after the field equations are derived (cf. also the explicit expressions in  \cite{Callan:1986bc}).
%
}
\begin{align} \label{cc:locaction}
S_\text{loc} &= \sum_p S_\text{loc}^{(p)} = \sum_p\left(S_\text{DBI}^{(p)} + S_\text{CS}^{(p)}\right)
\end{align}
with
\begin{align}
S_\text{DBI}^{(p)} = \mp \mu_p \int \star_{p+1} \e^{\tfrac{p-3}{4}\phi} \w \sigma(\delta_{9-p}), \qquad S_\text{CS}^{(p)} = \Bigg\{ \begin{split} & +\mu_p \int \langle C\wedge \e^{-B} \rangle_{p+1} \w \sigma(\delta_{9-p}) \\ & - \mu_p \int C_{p+1} \w \sigma(\delta_{9-p}) \end{split}, \label{cc:locaction2}
\end{align}
where the upper line is for D$p$-branes and the lower line for O$p$-planes, and $\mu_p > 0$ is the absolute value of the D$p$-brane/O$p$-plane charge. For $\overline{\textrm{D}p}$-branes and $\overline{\textrm{O}p}$-planes, the Chern-Simons terms would have the opposite sign. $\star_{p+1}$ is the Hodge operator on the $(p+1)$-dimensional world volume, $\Sigma$, of the source in question, and we define $\delta_{9-p}=\sigma(\star_{9-p} 1) \delta(\Sigma)$, where $\star_{9-p} 1$ is the $(9-p)$-dimensional volume form transverse to the source (defined such that $\star_{10} 1 = \star_{p+1} 1 \w \star_{9-p} 1$) and $\delta(\Sigma)$ is the delta distribution with support on $\Sigma$. We also use the polyform notation in \eqref{cc:locaction2}, \mbox{i. e.} $C= \sum_n C_{n-1}$ denotes the sum of all electric and magnetic RR potentials that appear in type IIA or type IIB supergravity, and $\e^{-B}$ is defined as a  power series of wedge products. The symbol $\langle\cdots \rangle_{p+1}$ denotes a projection to the form degree $p+1$, i.e., 
\begin{equation}
\langle C\w \e^{-B}\rangle_{p+1} = C_{p+1} - C_{p-1}\w B + \f{1}{2}C_{p-3}\w B\w B - \ldots
\end{equation}

Throughout this paper, we restrict ourselves to warped compactifications to $d \ge 4$ dimensions that preserve maximal symmetry in the non-compact $d$-dimensional spacetime. Accordingly, we only consider spacetime-filling sources extending in $p+1 \ge d$ dimensions. Furthermore, all fields are assumed to depend only on the internal coordinates $x^m$. The form fields are allowed to have legs in external directions only if they are spacetime-filling, in other words they have to be of rank $d$ or higher. All other form fields are purely internal. We assume a warped metric of the form
\begin{equation}
\d s_{10}^2 = g_{\mu\nu} \d x^\mu \d x^\nu + g_{mn} \d x^m \d x^n, \qquad g_{\mu\nu} = \e^{2A} \tilde g_{\mu\nu}, \label{metric}
\end{equation}
where $A$ is the warp factor and $\tilde g_{\mu\nu}$ is the unwarped $d$-dimensional metric corresponding to a Minkowski or (A)dS spacetime. We will also put a tilde on quantities such as Hodge operators, covariant derivatives or contractions of tensors if they are constructed using the unwarped metric instead of the warped one.

We now list the relevant equations of motion. The trace of the external Einstein equation reads
\begin{equation}
R_d = \frac{d}{2} \Big({\mathcal{L} - \sum_p\mathcal{L}^{(p)}_\text{CS}}\Big) + \frac{d}{4} \sum_n \e^{\tfrac{5-n}{2}\phi} |F_n^\ext|^2, \label{einsteinx}
\end{equation}
where $R_d = R_{\mu\nu} g^{\mu\nu}$ is the $d$-dimensional Ricci scalar and we denote the spacetime-filling RR field strengths by $F_n^\ext$. $\mathcal{L}$ is the Lagrangian including all bulk terms and the DBI and CS terms due to the localized sources, and $\mathcal{L}^{(p)}_\text{CS}$ are the CS parts of the source Lagrangian. For the warped metric \eqref{metric}, one finds
\begin{equation}
R_d = \frac{2d}{d-2} \e^{-2A} \Lambda - \e^{-dA} \tilde \nabla^2 \e^{dA}, \label{ricci-lambda}
\end{equation}
where $\Lambda$ is the $d$-dimensional cosmological constant.
Substituting this into \eqref{einsteinx} and integrating over ten-dimensional spacetime then yields
\begin{equation}
 \frac{8 v \mathcal{V}}{d-2} \Lambda = 2 \Big(S - \sum_p S^{(p)}_\text{CS}\Big) + \sum_n \int \star_{10}\, \e^{\tfrac{5-n}{2}\phi} |F_n^\ext|^2, \label{einstein-int}
\end{equation}
where we have introduced the volume factors
\begin{equation}
v = \int\ti\star_d 1,\qquad \mathcal{V} = \int \star_{10-d}\, \e^{(d-2)A}. \label{volumes}
\end{equation}

The Bianchi identities for the RR fields are
\begin{equation}
\d_{-H} F + j = 0, \label{bianchix}
\end{equation}
where $F= \sum_n F_n$ is the polyform containing the sum over all RR field strengths, $\d_{-H} = \d - H \w$ is the twisted exterior derivative, and $j$ is the polyform containing the sum over all source contributions of the different Bianchi identities, where $j = \sum_p \mu_p \langle\delta \w \e^{B}\rangle_{9-p}$ for D-branes and $j = - \sum_p \mu_p \delta_{9-p}$ for O-planes. Finally, we state the dilaton equation,
\begin{equation}
\nabla^2 \phi = - \frac{1}{2} \e^{-\phi} |H|^2 + \sum_n \frac{5-n}{8} \e^{\tfrac{5-n}{2}\phi} |F_n|^2 \pm \sum_p \frac{p-3}{4} \mu_p \e^{\tfrac{p-3}{4}\phi} \delta(\Sigma), \label{dilaton}
\end{equation}
and the equation of motion and Bianchi identity for $H$,
\begin{equation}
\d\left(\e^{-\phi}\star_{10} H\right) - \frac{1}{2} \left\langle F \w \sigma(F) \right\rangle_{8} = 0, \qquad \d H = 0. \label{h-eoms}
\end{equation}
\\


\section{The Cosmological Constant as a Sum of Source Terms}
\label{cc}

In this section, we will introduce two independent scaling symmetries satisfied by the action \eqref{cc:action} and use them to derive an expression for the cosmological constant $\Lambda$ in terms of the (on-shell evaluated) action of localized sources.

\subsection{Two Scaling Symmetries}\label{scaling}

It is known since the 1980s \cite{witten:1985xb} that the terms in the low energy effective action of string theory must satisfy simple scaling properties when the dilaton or equivalently the string coupling constant is scaled. This property is inherited from the simple coupling of the dilaton to the world sheet curvature in string perturbation theory and is manifest in the string frame of the 10D effective action. In Einstein frame, the scaling does not only affect the dilaton $\phi$, but also the metric $g_{MN}$ and the RR $(n-1)$-forms $C_{n-1}$ of the type II theories:
\begin{equation}
\e^{-\phi}\mapsto s\e^{-\phi},\quad g_{MN}\mapsto \sqrt{s} g_{MN},\quad
C_{n-1}\mapsto s C_{n-1},  \label{jhgskg}
\end{equation}
where $s$ is a scaling parameter. This then leads to
\begin{equation}
S^{(\chi)} \mapsto s^\chi S^{(\chi)}, 
\end{equation}
where $\chi$ is the Euler characteristic of the world sheet from which the contribution, $S^{(\chi)}$, to the effective action was derived. For a standard low energy effective action consisting of the classical two-derivative action for the bulk supergravity fields, $S_{\text{bulk}}$, and the lowest order action due to the presence of localized sources, $S_{\text{loc}}$, we then get
\begin{equation}\label{dilatonscaling}
S = S_{\text{bulk}} + S_{\text{loc}} \mapsto s^2 S_{\text{bulk}} + s S_{\text{loc}}.
\end{equation}
This can be verified using \eqref{jhgskg} in \eqref{cc:bulkaction} and \eqref{cc:locaction2} or simply by using the corresponding actions in the string frame. Thus, in absence of localized sources, the effect of \eqref{jhgskg} is to rescale the tree-level supergravity action by an overall factor $s^2$. The transformations \eqref{jhgskg} are then a symmetry of the theory, since they leave the equations of motion invariant.

A second scaling symmetry \cite{Burgess:1985zz}  can be obtained from the mass dimension of the fields, which can be determined from the fact that the effective action is a derivative expansion and has mass dimension zero. Using that the mass dimension of the coordinates is $-1$ and the mass dimension of a derivative is $+1$, one can count the number of derivatives of a given term in the action and the number of dimensions that are integrated over to determine the mass dimension of the fields. If one then scales the fields in the effective action according to their mass dimension but leaves the coordinates unscaled, one obtains a non-trivial scaling of the terms in the action. The corresponding scaling of the bosonic fields in type II string theory is\footnote{This symmetry is sometimes called \emph{Trombone} symmetry in the context of supergravity, see for example \cite{Cremmer:1997xj}. Note that in our conventions the exponent of $t$ in (\ref{gshogg}) actually corresponds to the length (i.e. the inverse mass) dimension of the field.}
\begin{equation}
g_{MN}\mapsto t^{-2} g_{MN},\quad C_{n-1}\mapsto t^{-(n-1)}C_{n-1},\quad 
B\mapsto t^{-2} B, \label{gshogg}
\end{equation}
where $t$ is another scaling parameter. This yields the following scaling of the terms in the low energy action
\begin{equation}
S_i^D \mapsto t^{i-D} S_i^D,
\end{equation}
where $D$ denotes the number of dimensions that are integrated over (usually $D=10$, but $D$ is less than ten for source terms) and $i$ denotes the number of derivatives of the terms involved.
For a two-derivative bulk action and zero-derivative source terms with $(p+1)$-dimensional world volume, we thus get
\begin{equation}\label{massscaling}
S = S_{\text{bulk}} + S_{\text{loc}} \mapsto t^{-8}S_{\text{bulk}} + 
\sum_p t^{-p-1}S^{(p)}_{\text{loc}},
\end{equation}
as can be verified using \eqref{gshogg} in \eqref{cc:bulkaction} and \eqref{cc:locaction2}. In absence of localized sources, the transformations \eqref{gshogg} are a symmetry, since they rescale the bulk action by an overall factor $t^{-8}$ and thus leave the equations of motion invariant. Together with \eqref{dilatonscaling}, this implies that the type II supergravity action at tree-level has two global scaling symmetries, which are explicitly broken by terms that are due to the presence of localized sources.

\subsection{The Method}\label{method}

As mentioned in the introduction, the above scaling symmetries can often be used to derive an expression for the cosmological constant $\Lambda$ in terms of the on-shell action of localized sources. In those cases where this is possible, $\Lambda$ is thus determined by the boundary conditions of some of the bulk supergravity fields at the positions of the sources and independent of the details of the dynamics in the bulk. We will argue below that this is possible for compactifications that involve at most NSNS $H$ flux and not more than one type of RR flux. This extends the recent work \cite{Burgess:2011rv} by an explicit evaluation of the general flux contribution and the use of a second scaling symmetry to gauge them away in the above-mentioned cases. The resulting simplified expression for the cosmological constant in terms of on-shell brane and O-plane actions can then be applied to several interesting type II flux compactifications, as we discuss in the remainder of the paper. 

The strategy for deriving our expression for $\Lambda$ is as follows. At first, the scaling symmetries are used to derive an expression for the action \eqref{cc:action} that holds on-shell. This on-shell expression can then be substituted into the integrated Einstein equation \eqref{einstein-int}, which, as we will show, eliminates the dependence of the equation on the bulk fields up to certain flux terms and yields the desired result for $\Lambda$. Before we discuss how to derive the on-shell action in the general case, let us at first review the basic principle \cite{Burgess:2011rv}  using a simple example. Consider an action $S[\psi_i]$ that depends on a number of fields $\psi_i$ and that satisfies a scaling symmetry,
\begin{equation}\label{scalingex}
S[\tau^{k_i}\psi_i] = \tau^k S[\psi_i],
\end{equation}
where the scaling parameter $\tau$ is a real number, and $k$ is assumed to be non-vanishing. We can then take the $\tau$ derivative of \eqref{scalingex} to obtain
\begin{equation}\label{scalingex2}
\int \sum_i k_i\tau^{k_{i}-1}\psi_i\de{S[\tau^{k_i}\psi_i]}{(\tau^{k_{i}}\psi_i)} = k\tau^{k-1}S[\psi_i],
\end{equation}
where we have 
written the result in terms of the the usual functional derivative
(which for derivative terms implicitly involves partial integrations). Evaluating the equation at $\tau=1$ and using the fact that the fields satisfy the equations of motion $\delta S[\psi_i]/ \delta\psi_i = 0$, we then find that the left-hand side of \eqref{scalingex2} vanishes and
\begin{equation}
S[\psi_i] = 0 \label{onshellaction}
\end{equation}
on-shell.

In deriving \eqref{onshellaction}, however, we made two simplifications that do in general not hold in the context of string compactifications. The right-hand side of the equation is therefore often more complicated than in this simple example. First, we assumed that all terms in the action $S[\psi_i]$ scale uniformly with $\tau$. When we identify $\tau$ with the scaling parameters $s$ and $t$ of the previous subsection, this is then not true in string theory when localized sources are included, as can be seen from \eqref{dilatonscaling} and \eqref{massscaling}.\footnote{The assumption would also break down if one includes, e.g., $\alpha^{\prime}$ or loop corrections.} Second, when we evaluated $\d S[\tau^{k_i}\psi_i]/\d \tau$ to arrive at \eqref{scalingex2}, we had to integrate by parts all those terms in $S[\tau^{k_i}\psi_i]$ that involve derivatives of $\psi_i$. In string theory, however, many compactifications involve the presence of non-trivial background fluxes. The corresponding NSNS and/or RR field strength(s) then have a non-exact part such that, globally, they cannot be written in terms of a gauge potential. Instead, their gauge potentials are only locally defined. Thus, total derivatives involving the NSNS or RR gauge potentials do not necessarily integrate to zero anymore but may involve non-trivial contributions from patches of different gauge charts, which would yield an extra contribution when one integrates by parts. When we repeat the  above calculation for the general action \eqref{cc:action}, we therefore expect that the right-hand side of \eqref{onshellaction} receives two contributions: one contribution due to the presence of localized sources and another one due to non-trivial background fluxes.

In order to account for the possibility of flux, we explicitly divide the NSNS and RR field strengths into a flux part, which is closed but not exact, and a fluctuation, which is exact and given in terms of a globally defined gauge potential. For $H$, we thus write
\begin{equation}
H= \d B + H^b, \label{def-nsns}
\end{equation}
where $H^b$ denotes the background flux and $B$ is the fluctuating globally defined NSNS potential. Since $H^b$ is closed, the Bianchi identity $\d H = 0$ is satisfied such that our definition is consistent.\footnote{We do not consider compactifications involving NS$5$-branes in this paper, i.e. the Bianchi identity for $H$ does not contain a source term.}

For the RR field strengths, separating off the non-exact part is more subtle. This is related to the fact that their Bianchi identities are more complicated and, in particular, that some of them receive contributions from localized sources. Since we only consider spacetime-filling sources in this paper, they enter the Bianchi identities as delta forms whose legs are always in some of the internal directions. Thus, a source term can only show up in the Bianchi identity for the purely internal part of the corresponding RR field strength. It is therefore convenient to split the polyform $F = \sum_n F_n$ into a part $F^\inte = \sum_n F^\inte_n$, which contains all RR field strengths that are purely internal and may have a source term in their Bianchi identity, and a part $F^\ext = \sum_n F^\ext_n$, which contains all RR field strengths that are spacetime-filling (and possibly also have legs in the internal part) and, accordingly, do not have a source term in their Bianchi identity,
\begin{equation}
F = F^\inte + F^\ext. \label{split-rrfields}
\end{equation}
For $F^\ext$, the Bianchi identities \eqref{bianchix} then simplify to
\begin{equation}
\d_{-H}F^\ext =0. \label{bi-rrfields}
\end{equation}
This allows us to make the ansatz
\begin{equation}
F^\ext = \d_{-H} C^\ext + \e^B\w F^b, \label{def-rrfields}
\end{equation}
where $F^b$ is a $d_{-H^b}$-closed but non-exact polyform containing the sum over the spacetime-filling background fluxes and $C^\ext$ is a polyform containing the sum over the spacetime-filling RR potentials. In a (maximally symmetric) type IIB compactification to $4$ dimensions, for example, we would have $F^b = F_5^b + F_7^b + F_9^b$ and $C^\ext = C_4^\ext + C_6^\ext + C_8^\ext$, since only forms of rank $4$ or higher would be allowed to be spacetime-filling. One can verify that \eqref{def-rrfields} solves the Bianchi identities \eqref{bi-rrfields} and is therefore a consistent ansatz for the field strengths $F^\ext$.

The Bianchi identities of the internal field strengths, $F^\inte$, however, may contain source terms such that these field strengths can in general not be written in a way similar to \eqref{def-rrfields} everywhere on the compact space. We will circumvent this problem in this paper by simply expressing, at the level of the equations of motion, $F^\inte$ in terms of their dual field strengths $F^\ext$, which then in turn can be expressed in terms of \eqref{def-rrfields}. If, for example, $F_3=F_3^\inte$ is internal, we can express it in terms of the spacetime-filling $F_7=F_7^\ext$ via the duality relation $F_3^\inte=- \e^{-\phi} \star_{10} F_7^\ext$ and then use \eqref{def-rrfields} to split $F_7^\ext$ into an exact and a non-exact part.\footnote{A subtlety occurs for $F_5$, which is self-dual, and $F_4$, which can have both internal and spacetime-filling components in compactifications to $4$ dimensions. In these cases, only the internal components $F_4^\text{int}$, $F_5^\text{int}$ can have a source term in the Bianchi identity. We therefore express those in terms of their duals $F_6^\text{ext}$, $F_5^\text{ext}$, which can in turn be written in terms of \eqref{def-rrfields}.}

Finally, let us note that, since we put the non-exact parts of the NSNS and RR field strengths into $H^b$ and $F^b$, we can assume that the gauge potentials $B$ and $C^\ext$ are globally defined. This implies that total derivatives involving $B$ and $C^\ext$ integrate to zero on a compact space, which will be used below. It should also be mentioned that, under the scalings \eqref{jhgskg} and \eqref{gshogg}, the flux terms $H^b$ and $F^b$ behave in the same way as the corresponding gauge potentials do. This follows from the fact that the mass dimension and the coupling to the dilaton is the same for the exact and the non-exact parts of the NSNS and RR field strengths.

\subsection{On-shell Action and Cosmological Constant}\label{onshellrel}

Let us now discuss how to derive the on-shell expression for the action \eqref{cc:action} that will later be used in the integrated Einstein equation \eqref{einstein-int} to obtain our result for $\Lambda$. Contrary to the simple example sketched in the previous subsection, the calculation is rather involved if one considers the general case including sources and fluxes. Let us therefore note that there is an alternative way to obtain our result, which only uses the equations of motion instead of exploiting the scaling symmetries. This second derivation may serve as a double-check of our results and is detailed in Appendix \ref{app_typeII}. In the following, we will continue to discuss the first method, using the scaling symmetries. The reader who is less interested in the technical details of the derivation may also jump directly to \eqref{y} and the subsequent discussion, where we present our result for $\Lambda$.

Let $\tau$ denote the scaling parameter, where $\tau$ equals $s$ if we consider the dilaton scaling \eqref{jhgskg} and $t$ in case of the mass scaling \eqref{gshogg}. Moreover, we will use primes to denote the $\tau$-transformed fields and the corresponding $\tau$-transformed action. Thus, if $\tau=s$, we have, for example, $g'_{MN} = \sqrt{s}g_{MN}$, and if $\tau=t$, we have $g'_{MN} = t^{-2}g_{MN}$.
According to \eqref{dilatonscaling} and \eqref{massscaling}, the action \eqref{cc:action} then scales as
\begin{equation}\label{scalingex3}
S' =  S'_\text{bulk} + S'_\text{loc} = \tau^k S_\text{bulk} + \sum_p \tau^{l_{p}} S^{(p)}_{\text{loc}},
\end{equation}
where $k=2$, $l_{p}=1$ for $\tau=s$ and $k=-8$, $l_{p}=-p-1$ for $\tau=t$. Taking the $\tau$ derivative and evaluating the equation at $\tau=1$, we find
\begin{equation}\label{scalingex4}
\left.{\frac{\d S'_\text{bulk}}{\d \tau}}\right|_{\tau=1} + \left.{\frac{\d S'_\text{loc}}{\d \tau}}\right|_{\tau=1} = k S_\text{bulk} + \sum_p l_{p} S^{(p)}_{\text{loc}}.
\end{equation}
We now proceed as in the simple example discussed in Section \ref{method}: we first evaluate the terms on the left-hand side of the equation and integrate by parts to express them in terms of a functional derivative of the action with respect to the fields. We then substitute the equations of motion to simplify the expressions.

The first term on the left-hand side of \eqref{scalingex4} yields\footnote{We define functional derivatives with respect to form fields, $A$,  such that $\delta S=\int \frac{\delta S}{\delta A} \wedge \delta A$.}
\begin{equation} \label{sbulkdiff}
\left.{\diff{S_\text{bulk}'}{\tau}}\right|_{\tau=1} = \int \left.{ \left[\de{S_\text{bulk}}{g_{MN}}\diff{g_{MN}'}{\tau} + \de{S_\text{bulk}}{\phi}\diff{\phi'}{\tau} + \de{S_\text{NSNS}}{H}\w \diff{H'}{\tau} + \left\langle\de{S_\text{RR}}{F}\w\diff{F'}{\tau}\right\rangle_{10}\right] }\right|_{\tau=1},
\end{equation}
where we have implicitly used partial integration to write the first two terms in the integrand as functional derivatives of $S_\text{bulk}$ with respect to the metric and the dilaton. These functional derivatives are equivalent to the variation of the bulk action, which will later allow us to use the equations of motion to simplify the expression. Similarly, we should also rewrite the remaining two terms in above equation as variations with respect to the NSNS and RR potentials. This is more involved since $H$ and $F$ may contain flux (cf. \eqref{def-nsns} and  \eqref{def-rrfields}), and so we will consider these terms separately later. Let us at first evaluate the $\d S'_\text{loc}/ \d \tau$ term in \eqref{scalingex4},
\begin{align}\label{locscaling}
\left.{\diff{S'_\text{loc}}{\tau}}\right|_{\tau=1} &= \int \left.{\left[\de{S_\text{loc}}{g_{MN}}\diff{g'_{MN}}{\tau} + \de{S_\text{loc}}{\phi}\diff{\phi'}{\tau} + \left\langle \de{S_\text{loc}}{C}\w\diff{C'}{\tau}\right\rangle_{10} + \de{S_\text{loc}}{B}\w\diff{B'}{\tau}\right]}\right|_{\tau=1} \notag \\ &=
\int \left.{\left[\de{S_\text{loc}}{g_{MN}}\diff{g'_{MN}}{\tau} + \de{S_\text{loc}}{\phi}\diff{\phi'}{\tau}\right]}\right|_{\tau=1} + \sum_p \diff{S_\text{CS}^{\prime (p)}}{\tau} \bigg|_{\tau=1}.
\end{align}
Since $S_\text{loc}$ does not depend on any field derivatives but only on the fields themselves, we did not have to integrate by parts here. We can now combine \eqref{sbulkdiff} and \eqref{locscaling} and use the equations of motion $\delta S / \delta g_{MN} = \delta S / \delta \phi = 0$ to obtain
\begin{align} \label{sscaling1}
\left.{\diff{S_\text{bulk}'}{\tau}}\right|_{\tau=1} + \left.{\diff{S'_\text{loc}}{\tau}}\right|_{\tau=1} &= \int \left.{\left[\de{S_\text{NSNS}}{H}\w \diff{H'}{\tau} + \left\langle\de{S_\text{RR}}{F}\w\diff{F'}{\tau} \right\rangle_{10}\right] }\right|_{\tau=1} + \sum_p \diff{S_\text{CS}^{\prime (p)}}{\tau} \bigg|_{\tau=1}.
\end{align}

The two terms involving $\delta H$ and $\delta F$ are evaluated as follows. Substituting \eqref{def-nsns} into the $\delta S_\text{NSNS}/\delta H$ term in \eqref{sscaling1}, we can integrate by parts to obtain 
\begin{align}
\int \left.{ \de{S_\text{NSNS}}{H}\w \diff{H'}{\tau}}\right|_{\tau=1}
 &= \int \left.{\left[\d \de{S_\text{NSNS}}{H}\w \diff{B'}{\tau} + \de{S_\text{NSNS}}{H}\w\diff{H'^b}{\tau}\right]}\right|_{\tau=1} \notag \\
 &=  \int \left.{\left[\de{S_\text{NSNS}}{B}\w\diff{B'}{\tau} + \de{S_\text{NSNS}}{H}\w\diff{H'^b}{\tau}\right]}\right|_{\tau=1}. \label{NSNSscaling}
\end{align}
The $\delta S_\text{RR}/\delta F$ term in \eqref{sscaling1} can be computed in a similar fashion but is more complicated due to the subtleties explained in Section \ref{method}. We first use \eqref{split-rrfields} and write
\begin{align} \label{frewrite0}
\int \left.{\left\langle\de{S_\text{RR}}{F}\w\diff{F'}{\tau}\right\rangle_{10}}\right|_{\tau=1} &= \int \left.{\left\langle\de{S_\text{RR}}{F^\ext}\w\diff{F'^\ext}{\tau} + \de{S_\text{RR}}{F^\inte}\w\diff{F'^\inte}{\tau}\right\rangle_{10}}\right|_{\tau=1}.
\end{align}
We now have to replace all RR field strengths $F_n^\inte$ by their dual field strengths $F_{10-n}^\ext$ in order to be able to write them in terms of the globally defined gauge potentials $C^\ext$ using \eqref{def-rrfields}, which in turn will allow us to integrate by parts in \eqref{frewrite0}. Using the duality relations \eqref{rrduality} as well as the scalings \eqref{jhgskg} and \eqref{gshogg}, we find for the two cases $\tau=s$ and $\tau=t$:
\begin{align}
\int \left.{\left\langle\de{S_\text{RR}}{F}\w\diff{F'}{s}\right\rangle_{10}}\right|_{s=1} &= \sum_n \int \left(\de{S_\text{RR}} {F_n^\ext}\w F_n^\ext + \de{S_\text{RR}}{F_n^\inte}\w F_n^\inte \right) \notag \\
&= \sum_n \int \left( \de{S_\text{RR}} {F_n^\ext}\w F_n^\ext - \de{S_\text{RR}}{F_{10-n}^\ext}\w F_{10-n}^\ext \right) \notag \\
&= 0, \displaybreak[3] \label{frewrite1} \\ 
\int \left.{\left\langle\de{S_\text{RR}}{F}\w\diff{F'}{t}\right\rangle_{10}}\right|_{t=1} &= \sum_n (1-n) \int \left(\de{S_\text{RR}} {F_n^\ext}\w F_n^\ext + \de{S_\text{RR}}{F_n^\inte}\w F_n^\inte \right) \notag \\
&= \sum_n (1-n) \int \left( \de{S_\text{RR}} {F_n^\ext}\w F_n^\ext - \de{S_\text{RR}}{F_{10-n}^\ext}\w F_{10-n}^\ext \right) \notag \\
&= \sum_n (10-2n) \int \de{S_\text{RR}}{F_n^\ext}\w F_n^\ext. \label{frewrite2}
\end{align}
These two expressions can now be rewritten in a way that will become convenient further below. In order to do so, we again exploit the scalings \eqref{jhgskg} and \eqref{gshogg} and make use of the identity $ {\delta S_\text{RR} / \delta F_n^\ext \w F_n^\ext} = - \frac{1}{2}\star_{10}\, \e^{(5-n)\phi/2} |F_n^\ext|^2$, which can be derived from \eqref{cc:bulkaction}. We thus find
\begin{align}
\int \left.{\left\langle\de{S_\text{RR}}{F}\w\diff{F'}{\tau}\right\rangle_{10}}\right|_{\tau=1} &= 2 \int \left.{\left\langle \de{S_\text{RR}}{F^\ext}\w \diff{F'^\ext}{\tau} \right\rangle_{10}}\right|_{\tau=1} - 2 k \int \left\langle \de{S_\text{RR}}{F^\ext}\w F^\ext \right\rangle_{10} \notag \\ & \quad - \frac{k}{2} \sum_n \int \star_{10}\, \e^{\tfrac{5-n}{2}\phi} |F_n^\ext|^2, \label{frewrite5}
\end{align}
where $k=2$ for $\tau=s$ and $k=-8$ for $\tau=t$ as in \eqref{scalingex3}.

We now integrate by parts on the right-hand side of equation (\ref{frewrite5}). Taking into account \eqref{def-nsns} and \eqref{def-rrfields},
this yields\footnote{The factor $\frac{1}{2}$ that appears when rewriting $\delta S_\text{RR}/\delta C^\ext$ in terms of $\delta S/\delta C^\ext$ and $\delta S_\text{loc}/\delta C^\ext$ is related to a subtlety regarding the variation of the CS action of the RR fields. One only obtains the correct equations of motion if one takes the coupling of the RR fields to the sources as being half the coupling that one would get from the ``naive'' variation of the action. One can think of this as being due to the fact that one half of $\sum_p S^{(p)}_\text{CS}$ represents an electric coupling of the RR fields to the sources, whereas the other half is due to a magnetic coupling of the dual RR fields to the sources. This subtlety is  
known in the literature and has, for example, been discussed in footnote 6 of \cite{Giddings:2001yu}.}
\begin{align} 
& \int \left.{ \left\langle\de{S_\text{RR}}{F^\ext} \w\diff{F'^\ext}{\tau}\right\rangle_{10} }\right|_{\tau=1} \notag \\
& = \int \left.{\left\langle\de{S_\text{RR}}{F^\ext}\w\left(\d_{-H}\diff{C'^\ext}{\tau} + \e^B\w\diff{F'^b}{\tau} - \diff{(\d B'+H'^b)}{\tau}\w C^\ext + \diff{B'}{\tau}\w\e^B\w F^b\right)\right\rangle_{10} }\right|_{\tau=1} \notag \\
& =  \left. \int\left\langle \de{S_\text{RR}}{C^\ext}\w\diff{C'^\ext}{\tau} + \de{S_\text{RR}}{F^\ext}\w \left({\e^B \w \diff{F'^b}{\tau} +  \de{F^\ext}{B} \w \diff{B'}{\tau}  + \de{F^\ext}{H} \w \diff{H'^b}{\tau}}\right)\right\rangle_{10}\right|_{\tau=1} \notag \\
& = \int \left. \left\langle \left(\de{S}{C^\ext}  - \frac{1}{2}\de{S_\text{loc}}{C^\ext}\right)\w\diff{C'^\ext}{\tau} + \f{1}{2}\de{S_\text{RR}}{B}\w\diff{B'}{\tau} \right.\right.\notag\\
&\qquad  + \left.\left.\de{S_\text{RR}}{F^\ext}\w \left({\e^B \w \diff{F'^b}{\tau} + \de{F^\ext}{H} \w \diff{H'^b}{\tau}}\right)\right\rangle_{10}\right|_{\tau=1} \notag \\
& = \int \left. \left\langle \left(\de{S}{C^\ext} - \frac{1}{2}\de{S_\text{loc}}{C^\ext}\right)\w\diff{C'^\ext}{\tau} + \f{1}{2}\left(\de{S}{B} - \de{S_\text{NSNS}}{B} - \de{S_\text{loc}}{B}\right)\w\diff{B'}{\tau}\right.\right.\notag\\
& \qquad + \left.\left.\de{S_\text{RR}}{F^\ext}\w\left(\e^B \w \diff{F'^b}{\tau} + \de{F^\ext}{H}\w\diff{H'^b}{\tau}\right)\right\rangle_{10}\right|_{\tau=1} ,
\end{align}
where we also used
\begin{equation}
2 \left\langle\de{S_\text{RR}}{F^\ext}\w\de{F^\ext}{B}\right\rangle_8 = \left\langle F^\ext \w \sigma(F^\inte)\right\rangle_8 - \de{S_\text{loc}}{B} = \de{S}{B} - \de{S_\text{NSNS}}{B} - \de{S_\text{loc}}{B} = \de{S_\text{RR}}{B},
\end{equation}
which can be derived using \eqref{cc:bulkaction}, \eqref{h-eoms}, \eqref{def-nsns} and \eqref{def-rrfields}. With the equations of motion, $\delta S/\delta C^\ext = \delta S/\delta B = 0$, one finally obtains 
\begin{align}\label{identity1}
\int \left. \left\langle\de{S_\text{RR}}{F^\ext} \w\diff{F'^\ext}{\tau}\right\rangle_{10} \right|_{\tau=1} &= -\frac{1}{2} \sum_p \diff{S^{\prime (p)}_\text{CS}}{\tau}\bigg|_{\tau=1} - \frac{1}{2} \int \left.{\de{S_\text{NSNS}}{B}\w\diff{B'}{\tau}}\right|_{\tau=1} \notag \\ & \quad + \int \left.{\left\langle \de{S_\text{RR}}{F^\ext}\w\left(\e^B \w \diff{F'^b}{\tau} + \de{F^\ext}{H}\w\diff{H'^b}{\tau}\right)\right\rangle_{10}}\right|_{\tau=1}.
\end{align}
Evaluating this for $\tau=s$ using \eqref{jhgskg} then also implies
\begin{align} \label{identity2}
\int \left\langle \de{S_\text{RR}}{F^\ext}\w F^\ext \right\rangle_{10} &= 
 -\frac{1}{2} \sum_p S^{(p)}_\text{CS} + \int \left\langle \de{S_\text{RR}}{F^\ext}\w \e^B \w F^b\right\rangle_{10}.
\end{align}
Substituting \eqref{identity1} and \eqref{identity2} into \eqref{frewrite5} then leads to
\begin{align}
\int \left.{\left\langle\de{S_\text{RR}}{F}\w\diff{F'}{\tau}\right\rangle_{10}}\right|_{\tau=1} =
& -\sum_p \diff{S^{\prime (p)}_\text{CS}}{\tau}\bigg|_{\tau=1} + k \sum_p S^{(p)}_\text{CS} - \int \left.\de{S_\text{NSNS}}{B}\w\diff{B'}{\tau}\right|_{\tau=1}\notag\\
& + 2\int \left.{\left\langle \de{S_\text{RR}}{F^\ext}\w\left(\e^B \w \diff{F'^b}{\tau} - k\ \e^B\w F^b + \de{F^\ext}{H}\w\diff{H'^b}{\tau}\right)\right\rangle_{10}}\right|_{\tau=1} \notag \\ 
&  - \frac{k}{2} \sum_n \int \star_{10}\, \e^{\tfrac{5-n}{2}\phi} |F_n^\ext|^2. \label{frewrite6}
\end{align}

Putting everything together, we now use \eqref{frewrite6} together with \eqref{NSNSscaling} in \eqref{sscaling1} to arrive at
\begin{align}\label{scalingresult}
\left[{\frac{\d S'_\text{bulk}}{\d \tau}} + {\frac{\d S'_\text{loc}}{\d \tau}}\right]_{\tau=1} =
&  k \sum_p S^{(p)}_\text{CS}  - \frac{k}{2} \sum_n \int \star_{10}\, \e^{\tfrac{5-n}{2}\phi} |F_n^\ext|^2 + \int\left.\de{S_\text{NSNS}}{H}\w\diff{H'^b}{\tau}\right|_{\tau=1}\notag \\
&  + 2\int \left.{\left\langle \de{S_\text{RR}}{F^\ext}\w\left(\e^B \w \diff{F'^b}{\tau} - k\ \e^B\w F^b + \de{F^\ext}{H}\w\diff{H'^b}{\tau}\right)\right\rangle_{10}}\right|_{\tau=1}.
\end{align}
Using \eqref{scalingex4} and the two scaling symmetries \eqref{jhgskg} and \eqref{gshogg} and evaluating the functional derivatives then leads to the two equations 
\begin{align}
2S_\text{bulk} + S_\text{loc} =
& 2\sum_p S_\text{CS}^{(p)}  - \sum_n \int \star_{10}\, \e^{\tfrac{5-n}{2}\phi} |F_n^\ext|^2\notag\\
& - \sum_n \int F^b_n \w\left\langle \e^B\w\sigma(F^\inte)\right\rangle_{10-n}, \label{onshell1} \\
-8S_\text{bulk} - \sum_p(p+1)S_\text{loc}^{(p)} =
& -8\sum_p S_\text{CS}^{(p)} +4 \sum_n \int \star_{10} \, \e^{\tfrac{5-n}{2}\phi} |F_n^\ext|^2\notag\\
& + \sum_n (9-n) \int F^b_n\w\left\langle \e^B \w \sigma(F^\inte)\right\rangle_{10-n} \notag \\ 
& - 2\int H^b\w \left( \e^{-\phi}\star_{10} H - \left\langle \sigma(F^\inte)\w C^\ext\right\rangle_7\right), \label{onshell2}
\end{align}
where $\sigma$ is the operator defined in \eqref{reversal}.
We can now linearly combine \eqref{onshell1} and \eqref{onshell2} introducing a free parameter $c$ and rearrange the source terms using $S = S_\text{bulk} + S_\text{loc}$ and $S_\text{loc}^{(p)} = S_\text{DBI}^{(p)} + S_\text{CS}^{(p)}$, which yields
\begin{align}
2S-2\sum_pS^{(p)}_\text{CS} + \sum_n \int \star_{10} \,
\e^{\frac{5-n}{2}\phi} |F_n^\ext|^2=
& \sum_p  \left(1+\f{p-3}{2}c\right) \left[S_{\text{DBI}}^{(p)}+S_\text{CS}^{(p)}\right]\notag\\
& - \sum_n \left(1+\f{n-5}{2}c\right) \int F^b_n \w\left\langle \e^B\w\sigma(F^\inte)\right\rangle_{10-n} \notag \\ 
& - c \int H^b\w \left( \e^{-\phi}\star_{10} H - \left\langle \sigma(F^\inte)\w C^\ext\right\rangle_7\right). \label{long}
\end{align}
Substituting this into the integrated Einstein equation \eqref{einstein-int} and collecting all contributions from background fluxes into a single term $\mathcal{F}(c)$, we find the result
\begin{equation}
\boxed{ \frac{8 v \mathcal{V}}{d-2} \Lambda = \sum_{p} \left(1+\f{p-3}{2}c\right) \left[S^{(p)}_\text{DBI} + S_\text{CS}^{(p)}\right] + \int \mathcal{F}(c)} \label{y}
\end{equation}
with the volume factors $v$ and $\mathcal{V}$ defined as in \eqref{volumes}. Note that all terms on the right-hand side of \eqref{y} contain an implicit factor of the external ``volume'' $v$ such that it cancels out in the equation, and $\Lambda$ does not depend on it. The flux term $\mathcal{F}(c)$ takes the form
\begin{align}
\mathcal{F} (c) &=  - \sum_{n \ge d} \left(1+\f{n-5}{2}c\right)F^b_{n}\w\left\langle\e^B\w \sigma(F^\inte)\right\rangle_{10-n} \notag \\ &\quad\, - cH^b\w\left( \e^{-\phi}\star_{10} H - \left\langle \sigma(F^\inte)\w C^\ext\right\rangle_7\right), \label{sigma}
\end{align}
where the summation range is determined by the fact that the background fluxes $F_n^b$ are spacetime-filling by definition and must therefore be of rank $d$ or higher (cf. the discussion in Section \ref{method}).

As stated earlier, the contribution of the flux term $\mathcal{F}(c)$ can often be gauged away in \eqref{y} by choosing an appropriate numerical value for the free parameter $c$. Up to an overall volume factor $\mathcal{V}$ (whose sign is known to be positive), $\Lambda$ is then completely determined by the on-shell actions of the localized sources that appear in the corresponding solution. If only one of the fluxes in \eqref{sigma} is non-zero, it is straightforward to see that $\mathcal{F}(c)$ can be set to zero, since then one can simply choose $c$ such that the $c$-dependent prefactor of the corresponding term vanishes in \eqref{sigma}.\footnote{$F_5$ flux is an exception, because it does not have a $c$-dependent prefactor in $\mathcal{F}(c)$ and can therefore not be gauged away in \eqref{y}. This is the reason for the existence of the Freund-Rubin solutions of type IIB supergravity on $\mathrm{AdS}_5 \times S^5$ \cite{Freund:1980xh}.} For a compactification with non-zero $H^b$, for example, 
 one would choose $c=0$, and, for a compactification with non-zero $F_7^b$, one would choose $c=-1$.

Even if the NSNS flux $H^b$ and one of the RR fluxes (other than $F_5^b$) are both non-zero, it is still often possible to find a $c$ such that $\mathcal{F}(c)$ vanishes. The reason is that the term multiplying $H^b$ in \eqref{sigma} is proportional to
\begin{equation}
\de{S_\text{NSNS}}{H}+2\left\langle\de{S_\text{RR}}{F^\ext} \w \de{F^\ext}{H}\right\rangle_7 = - \e^{-\phi}\star_{10} H + \left\langle \sigma (F^\inte)\w C^\ext\right\rangle_7. \label{gauge1}
\end{equation}
If the $H$ equation of motion implies that $\d \left[{\e^{-\phi}\star_{10} H - \left\langle \sigma (F^\inte)\w C^\ext\right\rangle_7}\right] = 0$, which is the case in many interesting examples, then we can write
\begin{equation}
- \e^{-\phi}\star_{10} H + \left\langle \sigma (F^\inte)\w C^\ext\right\rangle_7 = \omega_7, \label{gauge2}
\end{equation}
where $\omega_7$ is a closed but not necessarily exact $7$-form. Note that only a possible non-exact part of $\omega_7$ can contribute to \eqref{y} since any exact part of $\omega_7$ would reduce to zero when inserted into \eqref{sigma} and integrated over. If a gauge transformation of the RR potentials can be employed to cancel $\omega_7$ in \eqref{gauge2}, the term multiplying $H^b$ in \eqref{sigma} vanishes for any $c$, and we can choose the value for $c$ such that also the RR flux term in \eqref{sigma} vanishes. Consider, for example, a compactification of type IIA supergravity with non-zero $H^b$ and $F_0$. The non-trivial background fluxes appearing in \eqref{sigma} are then $H^b$ and $F^b_{10}$,
\begin{equation}
\mathcal{F} (c) =  - \left(1+\f{5}{2}c\right)F^b_{10}\w F_0 - cH^b\w\left( \e^{-\phi}\star_{10} H - \left\langle \sigma(F^\inte)\w C^\ext\right\rangle_7\right). \label{gauge4}
\end{equation}
Assuming that $\d \left[{\e^{-\phi}\star_{10} H - \left\langle \sigma (F^\inte)\w C^\ext\right\rangle_7}\right] = 0$ by the $H$ equation of motion, \eqref{gauge1} and \eqref{gauge2} now imply that the term multiplying $H^b$ can be canceled by a gauge transformation $C_7 \mapsto C_7 - \omega_7/F_0$. This is a valid gauge transformation that leaves all RR field strengths unchanged. In the new gauge, we then have $\e^{-\phi}\star_{10} H - \left\langle \sigma (F^\inte)\w C^\ext\right\rangle_7 =0$ such that \eqref{gauge4} reduces to $\mathcal{F}(c) = - (1+5c/2)F^b_{10} \w F_0$. We can therefore choose $c=-2/5$ so that $\mathcal{F} =0$.\footnote{Note that, even though $\mathcal{F}(c)$ is not gauge invariant, one can convince oneself that the full expression for $\Lambda$ in \eqref{y} is gauge invariant.}

In presence of more than one type of RR flux, this reasoning does not work anymore, since it is then not possible to choose an appropriate $c$ such that each term in $\mathcal{F}(c)$ is set to zero individually. We may still be able to find a $c=c_0$ that solves the equation $\int \mathcal{F}(c_0) = 0$ such that $\int \mathcal{F}(c_0)$ vanishes as a whole, but the numerical value of $c_0$ then depends on the bulk fields that appear in \eqref{sigma}. This will in general not be useful, since it just has the effect of trading the explicit dependence of $\Lambda$ on the bulk dynamics for an implicit dependence hidden in the value of $c_0$. We will explain this in more detail in Section \ref{examples} where we discuss several examples for string compactifications in which $\mathcal{F}(c)$ can be set to zero and one counterexample in which it cannot be set to zero.

\subsection{Validity of the Supergravity Approximation}

Before we proceed with applying the above results to some explicit examples, a comment on their regime of validity is in order. In the vicinity of localized sources, field derivatives and the string coupling often blow up such that $\alpha^\prime$ and loop corrections can become large, making the reliability of the supergravity approximation questionable. Given that the right-hand side of \eqref{y} is evaluated directly at the positions of the sources, one might therefore wonder about the self-consistency of our expression for $\Lambda$.

In order to clarify the meaning of our result, it is important to recall that (\ref{y}) has been derived by using the two-derivative supergravity action (\ref{cc:action}), (\ref{cc:bulkaction}), (\ref{cc:locaction}), (\ref{cc:locaction2}). Within this theory, (\ref{y}) is an \emph{exact} expression that can serve as well as any other method for calculating the cosmological constant in the supergravity approximation. The only question now is what  happens to (\ref{y}) if one takes into account the various types of stringy corrections, because these may significantly affect the strong field region at the sources.
 
The answer to this question depends on how (\ref{y}) is used. If one reads it as an expression that calculates the cosmological constant in terms of the near-source behavior, one has to use the near-source behavior in the supergravity approximation and then gets the cosmological constant in the supergravity approximation. Let us, for simplicity, focus on the case with only one type of sources present in the compactification. We can then schematically write $\Lambda^{\text{class}}=\kappa S_\textrm{loc}^\text{class}$, where the superscript $^\text{class}$ denotes the values in the supergravity appoximation, and $\kappa$ is some constant. If classical supergravity provides a good approximation for the lower-dimensional effective theory, e.g. in the usual
regime of large volume and small string coupling, the full cosmological constant, $\Lambda^{\textrm{full}}$, is well-approximated by the lowest order expression, $\Lambda^{\textrm{full}}\approx \Lambda^\text{class}$, and one therefore also has $\Lambda^{\textrm{full}}\approx \kappa S_\textrm{loc}^\text{class}$. Note that this is true even when $S_\textrm{loc}^\text{class}$ is not a good approximation to $S_\textrm{loc}^{\textrm{full}}$. This is the way we will use (\ref{y}) in Section \ref{examples}.

In Section \ref{kklt}, on the other hand, we also use (\ref{y}) backwards, i.e. we extract information on the near-brane behavior in a setup where $\Lambda$ is known. Here it is important to stress that this will only give us information on $S_\textrm{loc}^\text{class}$, i.e. on the near brane behavior in the supergravity approximation. In particular, the singularity in the $H$ and $F_3$ energy density we find is a priori only a feature of the supergravity approximation, and our result just confirms the singularity exactly like other people have seen the singularity in the supergravity approximation \cite{McGuirk:2009xx,Bena:2009xk,Bena:2011hz,Bena:2011wh,Bena:2012bk}. Whether the singularity gets resolved by stringy effects can not be inferred from our argument and is beyond the scope of our work. The useful advantage of our method is that it shows that this singularity survives the full supergravity analysis and is not an artifact of the partial smearing or a linearization around the BPS background.
\\


\section{Examples}
\label{examples}

In this section, we discuss different solutions of type IIA and IIB supergravity that have appeared in the literature and show
how \eqref{y} can be evaluated in our framework to obtain an explicit expression for the cosmological constant.
\\

\subsection{The GKP Solutions}
\label{Ex:GKP}

Here we consider warped compactifications of type IIB supergravity to $4$-dimensional Minkowski space with $H$ flux and $F_3$ flux and the necessary sources for tadpole cancelation along the lines of 
\cite{Giddings:2001yu} (GKP) and related work \cite{Dasgupta:1999ss,Gukov:1999ya,Becker:1996gj,Greene:2000gh}. For simplicity, we specialize to models involving only O$3$-planes as sources. In \cite{Giddings:2001yu}, 
the authors also discussed models with D$7$-branes and O$7$-planes along with their F-theory description. 
The discussion of  models with $7$-branes in our framework is analogous albeit more lengthy. 

Following \cite{Giddings:2001yu}, we find that the non-vanishing fields must satisfy
\begin{align}
F_3 = - \e^{-\phi} \star_6 H, \qquad F_5 = -(1+\star_{10}) \e^{-4A} \star_6 \d \alpha, \qquad C_4^\textrm{ext} = \ti \star_4(\alpha + a), \qquad \alpha = \e^{4A}, \label{gkp_fields}
\end{align}
where the warp factor $A$ and the dilaton $\phi$ are functions on the compact space, and $a$ 
is an integration constant corresponding to a gauge transformation. Also note that 
$F_5 =\star_{10} F_5 = F_5^\textrm{int} + F_5^\textrm{ext}$ with $F_5^\textrm{ext} = \d C_4^\textrm{ext}$. 
The topologically non-trivial fluxes canceling the O$3$-tadpoles are $F_3$ flux and $H$ flux, so that 
the relevant fluxes appearing in the definition of $\mathcal{F}(c)$, given by \eqref{sigma}, are 
\begin{equation}
H^b\quad \text{and}\quad F_7^b,
\end{equation}
whereas all other terms in \eqref{sigma} vanish. Thus \eqref{sigma} reduces to
\begin{equation}
\mathcal{F}(c) = -c\,H^b\w\left[\e^{-\phi} \star_{10} H + F_3\w C_4^\textrm{ext}\right] + \left({1+c}\right) F_7^b\w F_3. \label{gkp_sigma}
\end{equation}
Using \eqref{gkp_fields}, we find that the first term can be put to zero by gauge fixing $a=0$.\footnote{Note 
that, although $\mathcal{F}(c)$ is not gauge-invariant, the full expression for the cosmological constant $\Lambda$ is, since it contains a term $C_4 \w \mu_3 \delta_6$ which 
changes such that the total $a$-dependence of $\Lambda$ cancels out as it should.} Furthermore, $F_3$ and $H$ are related by a special condition which
is given in \eqref{gkp_fields}. This condition can be shown to saturate a BPS-like bound and is equivalent to the ISD condition of the complex three-form field 
strength in the notation of \cite{Giddings:2001yu}. It follows from this condition that also the second term in \eqref{gkp_sigma} is zero, as can be checked:
\begin{eqnarray}
\int F_7^b \w F_3  &=& \int  \left(F_7 - \d C_6^\textrm{ext} + H\w C_4^\textrm{ext}\right) \w F_3\notag\\
&=& \int\left(F_7 \w F_3 + \e^{\phi}\star_6 F_3 \w(\ti\star_4 \e^{4A})\w F_3\right) = 0,
\end{eqnarray}
where in the last step we used that $F_7 = -\e^{\phi}\star_{10} F_3 = -\e^{\phi}\star_6 F_3\w \ti \star_4 \e^{4 A}$. Thus $\mathcal{F}(c)$ reduces to zero for any choice of $c$. This is expected in this model, since also the contribution of localized source terms to $\Lambda$ is independent of 
$c$ for sources with $p=3$.

We therefore find that \eqref{y} yields
\begin{equation}
\Lambda = \frac{1}{4v\mathcal{V}} \left({S^{(3)}_\textrm{DBI} + S^{(3)}_\textrm{CS}}\right).
\end{equation}
Spelling out the contributions from the O$3$-planes and using \eqref{gkp_fields} in \eqref{cc:locaction2}, we arrive at
\begin{equation}
\Lambda = \frac{1}{4v\mathcal{V}}\, \mu_3\! \int \left(\ti\star_4\e^{4A} - C_4^\textrm{ext}\right) \w \sigma(\delta_6) = \frac{1}{4\mathcal{V}}\, N_{\textrm{O}3}\, \mu_3\! \left({\e^{4A_0} - \alpha_0}\right),
\end{equation}
where $A_0,\alpha_0$ denote the values of $A,\alpha$ at the position of the O$3$-plane(s) and $\mu_3 > 0$ is the absolute value of the O$3$ charge. Since $\alpha=\e^{4A}$, 
the DBI and Chern-Simons parts of the source action cancel out such that
\begin{equation}
\Lambda = 0
\end{equation}
as expected.
\\

\subsection[$\overline{\textrm{D}6}$-branes on $\mathrm{AdS}_7 \times S^3$]{$\overline{\textrm{D}\boldsymbol{6}}$-branes on $\boldsymbol{\mathrm{AdS}_7 \times S^3}$}
\label{d6}

Let us now consider type IIA supergravity with $\overline{\textrm{D}6}$-branes on $\mathrm{AdS}_7 \times S^3$, i.e. the setup 
studied in \cite{Blaback:2011nz,Blaback:2011pn,Bena:2012tx}.\footnote{Note that, unlike in the scenario considered in \cite{Kachru:2003aw}, the anti-branes are here not added to uplift an existing AdS solution to dS, but to cancel the tadpole and guarantee the existence of an AdS solution in the first place.} While a smeared solution can be constructed explicitly 
for this setup, it was argued in \cite{Blaback:2011nz,Blaback:2011pn} that in the supergravity approximation a solution with fully localized branes, 
if existent at all, necessarily yields a singularity in the energy density of the $H$ flux at the location of the $\overline{\textrm{D}6}$-branes. As we will see below,
it is rather straightforward to reproduce this result in our framework.

It was shown in \cite{Blaback:2011nz} that the non-vanishing fields in this setup must satisfy the ansatz
\begin{align}
F_0 = \textrm{const.}, \qquad H = \alpha F_0 \e^{\phi-7A} \star_3 1, 
\qquad F_2 = \e^{-3/2 \phi - 7A} \star_3 \d \alpha, \qquad C_7^\textrm{ext} = \tilde \star_7 (\alpha + a), \label{ads7_fields}
\end{align}
where the warp factor $A$, the dilaton $\phi$ and $\alpha$ are functions on the internal space, and $a$ is an integration constant related to a gauge freedom. 
The tadpole for the $\overline{\textrm{D}6}$-branes is canceled by a non-zero $H$ flux on the $3$-sphere and a non-zero Romans mass, i.e. $F_0$ ``flux''. 
The relevant fluxes appearing in $\mathcal{F}(c)$ are therefore
\begin{equation}
H^b\quad \text{and}\quad F^b_{10},
\end{equation}
and \eqref{sigma} reduces to
\begin{equation}
\mathcal{F} (c) =  -c\,H^b\w\left[\e^{-\phi} \star_{10} H - F_0\w C_7^\textrm{ext}\right] - \left({1+\frac{5}{2}c}\right) F_{10}^b\w F_0. \label{ads7_sigma}
\end{equation}
Using \eqref{ads7_fields}, one can see that the first term vanishes by a convenient gauge choice, $a=0$. We are then left with the second 
term which can be set to zero choosing $c=-\frac{2}{5}$.

We can now substitute this into \eqref{y} to find
\begin{equation}
\Lambda = \frac{1}{4v\mathcal{V}} \left({S^{(6)}_\textrm{DBI} + S^{(6)}_\textrm{CS}}\right).
\end{equation}
Spelling out the contributions of the $\overline{\textrm{D}6}$-branes and using \eqref{ads7_fields} then yields
\begin{equation}
\Lambda = \frac{1}{4v\mathcal{V}}\, \mu_6\! \int \left(- \ti\star_7\e^{3/4 \phi+7A} - C_7^\textrm{ext} \right) \w \sigma(\delta_3) =  
- \frac{1}{4\mathcal{V}}\, N_{\overline{\textrm{D}6}}\, \mu_6\! \left({ \e^{3/4 \phi_0+7A_0} + \alpha_0}\right), \label{d6-lambda}
\end{equation}
where $A_0,\alpha_0,\phi_0$ denote the values of $A,\alpha,\phi$ at the brane position and $\mu_6>0$ is the absolute value of the $\overline{\textrm{D}6}$ charge. Assuming that at leading order in the distance $r$ to the brane, 
the dilaton and the warp factor diverge as they would in flat space \cite{Janssen:1999sa}, 
\begin{equation}
\e^{2A} \sim r^{1/8}, \qquad \e^{\phi} \sim r^{3/4}, \label{wddiv}
\end{equation}
it is straightforward to show that the first term in \eqref{d6-lambda} (which comes from the DBI part of the brane action) is actually zero.
That this assumption is correct was explicitly proven in the analysis carried out in \cite{Blaback:2011pn}.

The cosmological constant is therefore exclusively determined by $\alpha_0$:
\begin{equation}
\Lambda \sim - \mu_6 \alpha_0.
\end{equation}
Since $\Lambda$ is negative, it then follows that $\alpha$ has to be non-zero and positive at the source. Together with \eqref{wddiv}, 
this implies that near the source the energy density of the $H$ flux diverges like the inverse of the warp factor,
\begin{equation}
\e^{-\phi} |H|^2 = \alpha^2\e^{-14A} \e^{\phi} F_0^2 \sim \e^{-2A}.
\end{equation}
This is consistent with the result found in \cite{Blaback:2011nz,Blaback:2011pn} by other methods, where it was also argued that 
finite $\alpha_0$ implies a singular energy density of the $H$ flux.
As we will show in Section \ref{kklt}, a similar argument holds for meta-stable de Sitter vacua that are obtained by placing $\overline{\textrm{D}3}$-branes on the Klebanov-Strassler throat embedded into a compact space. Under a few assumptions we will discuss in detail, one would find a singularity similar to the one observed in the $\overline{\textrm{D}6}$ model.
\\

\subsection[$\mathrm{SU}(3)$-structure Manifolds with O$6$-planes]{$\boldsymbol{\mathrm{SU}(3)}$-structure Manifolds with O$\boldsymbol{6}$-planes}

Here we discuss a particular model of compactifications of type IIA supergravity on $\mathrm{SU}(3)$-structure manifolds that was studied in \cite{Caviezel:2008tf}, 
namely O$6$-planes on $\mathrm{dS}_4 \times \mathrm{SU}(2) \times \mathrm{SU}(2)$ (see also \cite{Danielsson:2011au} for more examples of this type). This setup allows 
(unstable) critical points with positive $\Lambda$.

According to \cite{Caviezel:2008tf}, the form fields satisfy
\begin{align}
F_0 = m, \qquad F_2 = m^i Y_i^{(2-)}, \qquad H = p \left({Y_1^{(3-)}+Y_2^{(3-)}-Y_3^{(3-)}+Y_4^{(3-)}}\right),
\end{align}
where $Y_i^{(2-)},Y_i^{(3-)}$ are certain $2$-forms and $3$-forms, respectively, and $m,m^i,p$ are constant coefficients that are not relevant for the following discussion.
The tadpole generated by the O$6$-planes is canceled by non-zero $H$ and $F_0$ flux. However, while there is a non-trivial field strength $F_2$ (induced by the presence 
of the O$6$-planes), there is no topological $F_2$ flux, since it is not allowed by the cohomology of $\mathrm{SU}(2) \times \mathrm{SU}(2)$. For the same reason, $F_8^b=0$, and the 
non-zero background fluxes appearing in $\mathcal{F}(c)$ are
\begin{equation}
H^b\quad \text{and}\quad F^b_{10}.
\end{equation}
Considering \eqref{sigma} for this setup, we thus find
\begin{equation}
\mathcal{F}(c) = -c\,H^b\w\left[\e^{-\phi} \star_{10} H - F_0\w C_7^\textrm{ext}\right] - \left({1+\frac{5}{2}c}\right) F^b_{10} \w F_0. \label{su3_sigma}
\end{equation}
As discussed in Section \ref{onshellrel}, the $H$ equation of motion
\begin{equation}
\d \left[\e^{-\phi} \star_{10} H - F_0\w C_7^\textrm{ext}\right] = 0
\end{equation}
implies that we can choose a gauge for $C_7^\textrm{ext}$ such that the first term on the right-hand side of \eqref{su3_sigma} vanishes. The second term can be set to zero by choosing $c=-\frac{2}{5}$.

Evaluating \eqref{y}, we therefore find that the cosmological constant is given by
\begin{equation}
\Lambda = \frac{1}{10 v \mathcal{V}} \left({S^{(6)}_\textrm{DBI} + S^{(6)}_\textrm{CS}}\right) = \frac{1}{10 v \mathcal{V}}\, \mu_6\! \int \left({ \e^{3/4 \phi} \, \star_4 1 \w \star_3 1 - C_7^\textrm{ext}}\right) \w \sigma(\delta_3), \label{lambda-su3}
\end{equation}
where the right hand side should be understood as a sum over the various O$6$-plane terms, and 
$\mu_6 > 0$ is the absolute value of the O$6$ charge. In \cite{Caviezel:2008tf}, the setup was considered in the smeared limit, where the delta forms $\delta_3$ are replaced by volume forms of the space transverse to the corresponding sources. If a localized version of this solution exists, \eqref{lambda-su3} would give a constraint on the possible field behavior at the O-planes.
\\

\subsection{The DGKT Solutions}

Finally, we look at type IIA supergravity compactified on $T^6/\Z^2_3$, which is an explicit example for the type IIA flux
compactifications considered in \cite{DeWolfe:2005uu,Acharya:2006ne}.\footnote{As discussed in \cite{Acharya:2006ne}, the sources are smeared in order to obtain a solution. The discussion whether a corresponding localized solution exists or how it differs from the smeared solution \cite{Acharya:2006ne,Banks:2006hg,Saracco:2012wc,McOrist:2012yc,Douglas:2010rt,Blaback:2010sj,Blaback:2011nz} does not concern us here. We only consider this model to give an example of a solution where many fluxes are turned on.} In order to stabilize the moduli, the model 
requires the presence of NSNS flux as well as several RR fluxes of different ranks. As discussed in Section \ref{onshellrel}, 
it is therefore a counterexample, where it is in general not possible to set the flux-dependent terms in \eqref{y} to zero and write $\Lambda$ as a sum of localized source terms only.

The NSNS and RR field strengths in this model are given by 
\begin{equation}
H^b = -p \beta_0,\quad F_0 = m_0, \quad F_2=0, \quad F_4 = F_4^\inte + F_4^\ext = e_i \ti\omega^i + \star_{4}\, e_0,
\end{equation}
where $p,m_0,e_0,e_i$ are numbers, $\beta_0$ is an odd $3$-form and $\ti\omega^i$ are even $4$-forms under the orientifold involution.\footnote{Note that the spacetime-filling part of $F_4$, which is given by $F_4^\ext$, is treated as internal $F_6$ in the conventions of \cite{DeWolfe:2005uu}.} The non-trivial fluxes appearing in \eqref{sigma} are thus
\begin{equation}
H^b,\quad F^b_{10},\quad F^b_{6}\quad \text{and}\quad F^b_{4}
\end{equation}
such that
\begin{align}
\mathcal{F}(c) &= - c\,H^b\w \left[ \e^{-\phi}\star_{10} H - F_0\w C_7^\textrm{ext}\right] - \left(1 + \f{5}{2}c\right) F_{10}^b\w F_0 \notag \\
&\quad - \left(1+\f{1}{2}c\right)F_6^b\w F_4^\inte + \left(1 - \f{1}{2}c\right) F_4^b \w F_6^\inte, \label{sigma_dgkt}
\end{align}
where we used that the fluctuation $B$ is zero on-shell. The first term on the right-hand side can be made to vanish by choosing a gauge for the $C_7^\textrm{ext}$ field. Since the other terms do in general not vanish,
however, we cannot choose $c$ such that all of them are set to zero simultaneously.

As pointed out in Section \ref{onshellrel}, we can still solve the equation $\int \mathcal{F}(c) = 0$ for some $c=c_0$ (unless its $c$-dependence coincidentally
cancels out on-shell) and use it in \eqref{y} to arrive at an expression for $\Lambda$ which formally only depends on source terms,
\begin{equation}
\Lambda = \frac{2+3c_0}{8v \mathcal{V}} \left({S^{(6)}_\textrm{DBI} + S^{(6)}_\textrm{CS}}\right).
\end{equation}
However, the resulting numerical value for $c_0$ then implicitly depends on the bulk fields appearing in $\mathcal{F}(c)$. It is therefore hard
to approximate its numerical value or even its sign in compactification scenarios with more than one type of RR flux, unless the full solution
is already known (as in the present example). This is contrary to the previous examples, where $c$ could be fixed to a known number such that,
up to a volume factor, $\Lambda$ was completely determined by the boundary conditions of the fields in the near-source region.
\\


\section{Singular $\overline{\textrm{D}\boldsymbol{3}}$-branes in the Klebanov-Strassler Throat}
\label{kklt}

In this section, we discuss to what extent our previous
results can be applied to meta-stable de Sitter vacua in type IIB
string theory obtained by placing $\overline{\textrm{D}3}$-branes 
at the tip of a warped throat geometry
 along the lines of \cite{Kachru:2003aw}. We spell out and discuss the assumptions under which one can give a simple topological
argument for a singularity in the energy density of  $H$ and $F_3$ due to the brane backreaction.
\\

\subsection{Ansatz}

Following \cite{Kachru:2003aw}, we consider type IIB no-scale Minkowski solutions obtained by embedding the 
Klebanov-Strassler solution \cite{Klebanov:2000hb} into a compact setting \cite{Giddings:2001yu}. In order 
to stabilize the geometric moduli, we also include non-perturbative effects which may come from Euclidean D$3$-brane instantons or gaugino condensation. 
The resulting supersymmetric AdS vacuum is then uplifted to a meta-stable de Sitter vacuum by putting a small number of $\overline{\textrm{D}3}$-branes 
at the tip of the Klebanov-Strassler throat \cite{Kachru:2002gs,Kachru:2003aw}.

In order to apply the results of Section \ref{cc} to this scenario, we split the total cosmological constant into a part, $\Lambda^\textrm{class}$, 
which is due to the classical equations of motion and given by evaluating \eqref{y} at the solution, and the rest, $\Lambda^\textrm{np}$, 
which contains all corrections from non-perturbative effects that are not captured by the classical computation, i.e., we write
\begin{equation}
\Lambda = \Lambda^\textrm{class} + \Lambda^\textrm{np}. \label{lambda_split}
\end{equation}
Let us now discuss the explicit form of $\Lambda^\textrm{class}$ in the present setup. For simplicity, we will restrict ourselves to the case, 
where the no-scale solutions of \cite{Giddings:2001yu} are realized in a model with O$3$-planes, and the non-perturbative effects 
come from Euclidean D$3$-brane instantons. In \cite{Giddings:2001yu}, also orientifold limits of F-theory compactifications involving D$7$-branes and O$7$-planes 
are discussed. We checked that it is also possible to study such models in our framework, but the discussion becomes more involved, since the presence 
of these sources induces a non-trivial $F_1$ field strength. 

Our ansatz for the different fields thus reads\footnote{If one no longer assumes the BPS condition of Section \ref{Ex:GKP}, the function $\alpha$ need not be related to the warp factor, and $X_3$ may be non-vanishing.}
\begin{equation}
C_4^\textrm{ext} = \tilde \star_4 (\alpha + a), \quad F_5 = - (1+\star_{10}) \e^{-4A} \star_6 \d \alpha, \quad H = \e^{\phi-4A} \star_6 \left({ \alpha F_3 + X_3 }\right), \quad F_1 = 0, \label{kklt_fields}
\end{equation}
where $A,\alpha,\phi$ are functions on the internal space, $a$ is an integration constant corresponding to a gauge freedom, and $X_3$ is an a priori unknown $3$-form satisfying $\d X_3=0$. One can check that this ansatz follows from the form equations of motion and the requirement that the non-compact part of space-time be maximally symmetric, if only sources with $p=3$ are present.

As in the examples discussed in Section \ref{examples}, the flux-dependent terms $\mathcal{F}(c)$ in \eqref{y} can now be simplified 
by a convenient choice of the parameter $c$. To see this recall that the relevant fluxes in the present case are
\begin{equation}
H^b\quad\text{and}\quad F_7^b
\end{equation}
and thus \eqref{sigma}  reduces to
\begin{align}
\mathcal{F} (c) = & - c\, H^b\w\left[\e^{-\phi} \star_{10} H + F_3\w C_4^\ext\right] + \left({1+c}\right) F^b_7 \w F_3. \label{sigma-kklt}
\end{align}

Using \eqref{kklt_fields}, we find that the first expression on the right-hand side of \eqref{sigma-kklt} cancels out for $a=0$ except for a term $\sim\! X_3$. The second term in \eqref{sigma-kklt} can be set to zero by the choice $c=-1$, yielding\footnote{To be precise, one finds that the integrated dilaton equation implies
$ - \int H^b\w\left[\e^{-\phi} \star_{10} H + F_3\w C_4^\ext\right] + \int F^b_7 \w F_3 = 0$ in absence of sources with $p \neq 3$, such that 
$\int \mathcal{F} (c) = - \int \tilde \star_4 1 \w H^b \w X_3$ actually holds for any choice of $c$. This is consistent with the fact that also the source part of 
\eqref{y} is independent of $c$ for $p=3$. Thus the value of $\Lambda^\textrm{class}$ is uniquely determined by \eqref{y} as it should be.}
\begin{equation}
\mathcal{F} (-1) = - \tilde \star_4 1 \w H^b\w X_3. \label{sigma-kklt2}
\end{equation}
We will argue below that, upon a certain choice for the UV boundary conditions of the three-form field strengths, the integral of \eqref{sigma-kklt2} gives a contribution to the cosmological constant in  \eqref{y} that is negligible compared to the contribution from the anti-D3-brane source terms.

Keeping the flux term for the moment, we can substitute \eqref{sigma-kklt2} into \eqref{y} and  write
\begin{align}
\Lambda^\textrm{class} &= \frac{1}{4v\mathcal{V}} \left({S^{(3)}_\textrm{DBI} + S^{(3)}_\textrm{CS}}\right) + \frac{1}{4v\mathcal{V}} \int \mathcal{F} (-1) \notag \\
&= \frac{1}{4v\mathcal{V}}\, \mu_3\! \int \left( -\ti\star_4\e^{4A} - C_4^\textrm{ext}\right) \w \sigma\big(\delta^{(\overline{\textrm{D}3})}_6\big) + \frac{1}{16v\mathcal{V}}\, \mu_3\! \int \left(\ti\star_4\e^{4A} - C_4^\textrm{ext} \right) \w \sigma\big(\delta^{(\textrm{O}3)}_6\big) \notag \\ & \quad - \frac{1}{4v\mathcal{V}} \int \tilde \star_4 1 \w H^b \w X_3,
\end{align}
where we have spelled out the contributions of the localized sources. Note that the O$3$-plane charge is $\frac{1}{4}$ of the $\overline{\textrm{D}3}$-brane charge $\mu_3$, where $\mu_3 >0$ in our conventions. Evaluating the above equation, we find that the total cosmological constant \eqref{lambda_split} is given by
\begin{equation}
\Lambda = - \frac{1}{4\mathcal{V}}\, N_{\overline{\textrm{D}3}}\, \mu_3\! \left({\e^{4A_0} + \alpha_0}\right) + \frac{1}{16\mathcal{V}}\, N_{\textrm{O}3}\, \mu_3\! \left({ \e^{4A_*} - \alpha_*}\right) - \frac{1}{4\mathcal{V}} \int_{\mathcal{M}^{(6)}} H^b \w X_3 + \Lambda^\textrm{np}, \label{lambda_kklt}
\end{equation}
where $A_0,\alpha_0$ and $A_*,\alpha_*$ denote the values of $A,\alpha$ at the positions of the $\overline{\textrm{D}3}$-branes and O$3$-planes, respectively.
\\

\subsection{The Argument}

Our goal is now to evaluate \eqref{lambda_kklt} and relate it to the near-tip behavior of the energy density of the $H$ flux. In order to do so, we make the following assumptions.
\begin{enumerate}
\item {\bf Topological flux.}
In the region of the conifold, $F_3$ carries a non-trivial topological flux along the directions of a $3$-cycle called the A cycle, $H$ carries a topological flux along the directions of the dual $3$-cycle called the B cycle, and all other components of $H$ and $F_3$ are exact.  This assumption is due to the fact that the deformed conifold is topologically a cone over $S^2 \times S^3$, where the deformation has the effect of replacing the singular apex of the conifold by a finite $S^3$ (see e.g. \cite{Candelas:1989js,Minasian:1999tt}). The deformed conifold therefore has a non-trivial compact $3$-cycle along the $S^3$ (the A cycle) and a dual, non-compact $3$-cycle (the B cycle). We will assume that also in our compact setting the relevant cycles threaded by topological flux are the A cycle and the B cycle, at least in the region of the conifold. Following the literature \cite{Klebanov:2000hb}, we then place $F_3$ flux along the A cycle and $H$ flux along the B cycle. On general compact manifolds, there may of course exist additional cycles that are threaded by flux. We will assume, however, that such additional topologically non-trivial terms in $F_3$ and $H$ only become relevant deep in the UV, i.e., far away from the anti-D3-brane. 

\item {\bf IR boundary conditions.} The $\overline{\textrm{D}3}$-brane locally deforms the geometry as it would do in flat space. 
This implies in particular that the warp factor goes to zero in the vicinity of the $\overline{\textrm{D}3}$-brane as it usually does,
\begin{equation}
\e^{2A} \to 0. \label{wftozero}
\end{equation}
It also implies that we can locally approximate the internal geometry by 
\begin{equation}
g_{mn} \approx \e^{-2A} \tilde g_{mn}
\end{equation}
at leading order in an expansion around the distance $r$ to the brane, with $\tilde g_{mn}$  regular (in suitable coordinates). 

This is a standard assumption discussed recently e.g. in \cite{Bena:2012bk,Bena:2012vz} for the case of partially smeared $\overline{\textrm{D}3}$-branes.
In an analogous setting, it was verified explicitly in \cite{Blaback:2011pn} for the toy model with $\overline{\textrm{D}6}$-branes discussed in Section \ref{d6}, 
where both the warp factor and 
the internal metric indeed diverge exactly as they would do in the corresponding flat space solution \cite{Janssen:1999sa} at 
leading order in the distance parameter $r$. It would be interesting to carry out a similar derivation as in
\cite{Blaback:2011pn} also for the $\overline{\textrm{D}3}$-branes considered here, but this is  beyond the scope of the present paper 
(see also \cite{Bena:2012bk,Bena:2012vz} for an analogous discussion of partially smeared $\overline{\textrm{D}3}$-branes 
in the non-compact Klebanov-Strassler solution).

In order that the unperturbed deformed conifold metric $\tilde g_{mn}$ shrinks smoothly at the tip, we furthermore expect that the energy density of $F_3$ along the A cycle contracted with $\tilde g_{mn}$ does not vanish at the tip:
\begin{equation}
\e^\phi |\tilde F^A_3|^2 \neq 0, \label{finitef3}
\end{equation}
where the superscript denotes the component of $F_3$ along the A cycle.\footnote{This is not to be confused with the notation of \cite{Massai:2012jn,Bena:2012vz}, where the superscript in $F_3^A$ is an index running over all components of $F_3$.}
This is motivated by the fact that the energy density of  $F_3^A$ is non-vanishing and prevents the A cycle from collapsing at the tip of the deformed conifold before the perturbation by the $\overline{\textrm{D}3}$-branes \cite{Klebanov:2000hb}. Using the results of \cite{Bena:2012vz}, one can verify that \eqref{finitef3} indeed holds for the case of partially smeared $\overline{\textrm{D}3}$-branes.

\item {\bf UV boundary conditions.} The boundary conditions for the O$3$-planes in the UV far away from the $\overline{\textrm{D}3}$-branes are approximately the standard BPS 
boundary conditions,
\begin{equation}
\alpha_* \approx \e^{4A_*},\label{distortedrelation}
\end{equation}
up to small corrections such that the O$3$-plane term in \eqref{lambda_kklt} is negligible  compared to the other terms. To justify this, recall that in the GKP setup
without the $\overline{\textrm{D}3}$-branes this is the usual BPS behavior that does not lead to a contribution to the cosmological constant. 
When a large flux background with a large number of O$3$-planes of this type is then perturbed by a small number of $\overline{\textrm{D}3}$-branes at the tip of a warped throat,
the $\overline{\textrm{D}3}$-branes will give a small direct contribution to the cosmological constant due to their tree-level brane action (see below). One might however wonder whether the
$\overline{\textrm{D}3}$-brane backreaction on the geometry and the fields could also distort the relation (\ref{distortedrelation}) near the O$3$-planes, such that now also the O$3$-planes
would 
contribute significantly to the vacuum energy. However, this backreaction effect would be of higher order in the small perturbation from the redshifted
$\overline{\textrm{D}3}$-branes and should thus be negligible compared to the
direct 
contribution from the $\overline{\textrm{D}3}$-brane
source terms. This  is analogous to the usual assumption of BPS asymptotics 
in the UV imposed in non-compact treatments of brane backreaction (e.g. \cite{Bena:2012tx,Bena:2012bk}). 
It would be an interesting extension to explicitly compute the boundary conditions at the O-planes, e.g. following the analysis in \cite{Blaback:2011pn}.

Similarly, we also assume that the three-form field strengths approach their unperturbed values and thus become ISD in the UV far away from the $\overline{\textrm{D}3}$-branes, which implies
\begin{equation}
X_3^\textrm{UV} \approx 0, \label{uv-isd}
\end{equation}
again up to corrections that are negligible in \eqref{lambda_kklt}.
One might again wonder whether a small deviation from the ISD condition in the UV due to the anti-brane backreaction might be relevant for the value of the cosmological constant. As discussed above, however, it would be very surprising if the effect of such a deviation far away from the $\overline{\textrm{D}3}$-branes would not be negligible compared to their direct effect in the IR, so that we will adopt \eqref{uv-isd} as a reasonable assumption.

\item {\bf Non-perturbative corrections.} Non-perturbative corrections to the effective potential (due to, e.g.,  Euclidean D$3$-branes or gaugino condensation on D7-branes) 
are captured by adding a \emph{negative} term to the overall cosmological constant, i.e.
\begin{equation}
\Lambda = \Lambda^\textrm{class} - |\Lambda^\textrm{np}|. 
\end{equation}
This assumption consists in fact of two parts: The first is that the non-perturbative effect gives, by itself, rise to a negative contribution to the vacuum energy, 
and the second is that
it does not significantly change the classical contributions. These assumptions are implicit in the construction of \cite{Kachru:2003aw}, where the non-perturbative effects
first make the vanishing cosmological constant of the GKP setup negative without significantly changing the classical background fluxes or the vevs and masses of the moduli that are stabilized by 
these fluxes (the complex structure moduli and the dilaton). 
Moreover, the subsequent de Sitter uplift due to $\overline{\textrm{D}3}$-branes is assumed to happen through their classical source terms only and
does in turn not significantly change the vevs and masses of the moduli that are stabilized by the non-perturbative effects (the K\"{a}hler moduli). 
There has also been some progress in describing the above effects from an
explicit 10D point of view \cite{Koerber:2007xk,Baumann:2010sx,Dymarsky:2010mf,Heidenreich:2010ad}. In \cite{Heidenreich:2010ad} it was argued 
that a non-vanishing gaugino bilinear $\langle\overline{\lambda}\lambda\rangle$ on D7-branes indeed leads to a negative contribution to the 
 4D spacetime curvature proportional to $|\langle\overline{\lambda}\lambda\rangle|^2$. On the other hand, the backreaction of this on the classical contribution 
 $\Lambda^{\textrm{class}}$ to the vacuum energy 
 would be only a higher order effect. Similar properties are expected for the non-perturbative corrections due to 
 Euclidean D3-brane instantons.

\item {\bf Cosmological constant.} The presence of the $\overline{\textrm{D}3}$-branes uplifts the solution to a meta-stable de Sitter vacuum such that the total cosmological constant of the solution is positive,
\begin{equation}
\Lambda > 0,
\end{equation}
as proposed in \cite{Kachru:2003aw}.
\end{enumerate}

If one makes the above assumptions 1. - 5., our ansatz \eqref{lambda_kklt} for the cosmological constant drastically simplifies. 

Let us at first discuss the flux term in \eqref{lambda_kklt}.
Since $X_3$ is closed by definition, we can make the ansatz
\begin{equation}
X_3 = \beta \omega_3^A + \d \omega_2 \label{bla}
\end{equation}
in the conifold region. Here $\beta$ is an unknown function of the internal coordinates, $\omega_2$ is a $2$-form, and $\omega_3^A$ is the harmonic $3$-form along the A cycle satisfying $\d \omega_3^A = 0$. We have split $X_3$ into a part, $\beta \omega_3^A$, along the A cycle, which can in general be non-exact, and a part, $\d\omega_2$, that is not necessarily along the A cycle and has to be exact.\footnote{Note that, assuming the presence of $F_3$ flux along the A cycle, $X_3$ is not allowed to have a non-exact component along the B cycle as follows from the $F_1$ equation $\e^{-\phi} H \w \star_{10} F_3 = 0$ and the ansatz for $H$ stated in \eqref{kklt_fields}.} Using $\d X_3 = \d \omega_3^A = 0$, we find from \eqref{bla} that
\begin{equation}
\d \beta \w \omega_3^A = 0,
\end{equation}
which implies that $\beta$ is only a function of the coordinates parametrizing the $S^3$ but constant over the remaining directions. We can therefore set $\beta = \beta^\textrm{UV}=0$ without loss of generality, where $\beta^\textrm{UV}$ denotes the value of $\beta$ in the UV region of the warped throat far away from the $\overline{\textrm{D}3}$-branes.

The flux term in \eqref{lambda_kklt} then simplifies as follows. Since, under assumption 1., $H$ only carries a flux along the B cycle in the conifold region, we find $H^b \w X_3 = H^b \w (\beta \omega_3^A + \d \omega_2) = H^b \w \beta^\textrm{UV} \omega_3^A -\d (H^b \w \omega_2)$. We can therefore write
\begin{equation}
\int_{\mathcal{M}^{(6)}} H^b \w X_3 = \int_{\mathcal{M}^{(6)}} H^b \w X^\textrm{UV}_3 =0\label{bla2}
\end{equation}
such that the integral is completely determined by the units of $H$ flux present in the compactification and the UV boundary conditions for the three-form field strengths but independent of the IR physics close to the $\overline{\textrm{D}3}$-branes.

Using \eqref{bla2} together with assumptions 2.- 4., we find that \eqref{lambda_kklt} reduces to 
\begin{equation}
\Lambda \approx - \frac{1}{4\mathcal{V}}\, N_{\overline{\textrm{D}3}}\, \mu_3\, \alpha_0 - |\Lambda^\textrm{np}|,
\end{equation}
up to negligible corrections. From assumption 5. it then follows that
\begin{equation}
- \frac{1}{4\mathcal{V}}\, N_{\overline{\textrm{D}3}}\, \mu_3\, \alpha_0 > |\Lambda^\textrm{np}|,
\end{equation}
which implies that $\alpha_0$ must be finite and negative.\footnote{Note that $\alpha$ must change its sign somewhere in between the BPS region around the O$3$-planes (where $\alpha \approx \e^{4A}$) and the tip of the throat (where $\alpha < 0$). In the toy model discussed in \cite{Blaback:2011nz}, a similar constraint was used to formulate a topological no-go theorem, which is rederived in our framework in Section \ref{d6}.}

It is straightforward to see that this yields a singular energy density of the $H$ flux in the region near the $\overline{\textrm{D}3}$-branes. 
As argued above, we can locally approximate the internal metric as $g_{mn} \approx \e^{-2A} \tilde g_{mn}$, where $\tilde g_{mn}$ is 
regular. Using \eqref{kklt_fields}, we can then write 
\begin{equation}
\e^{-\phi} |H|^2 = \e^{\phi-8A} |\alpha F_3 + X_3|^2 \ge \alpha^2\e^{-8A} \e^{\phi} |F^A_3|^2 \approx \alpha^2\e^{-2A} \e^{\phi} |\tilde F^A_3|^2
\end{equation}
in the near-brane region, where we have used that the component of $X_3$ along $F^A_3$ vanishes. Since $e^{\phi}|\tilde F_3^A|^2$ is expected to be non-zero at the tip of the conifold, it then follows from \eqref{wftozero} and $\alpha_0 \neq 0$ that the energy density of the $H$ flux at least diverges like the inverse of the warp factor,
\begin{equation}
\e^{-\phi} |H|^2 \sim \e^{-2A}.
\end{equation}
Assuming a regular dilaton\footnote{If the dilaton diverges at the brane even though it does not directly couple to it, $\e^{-\phi} |H|^2 $ would still diverge, but the dilaton equation would not necessarily imply that $\e^{\phi} |F_3|^2$ also diverges.}, the   dilaton equation \eqref{dilaton} furthermore implies that the divergence in the energy density of $H$ must be canceled by a divergent term in the energy density of $F_3$. We thus find that the energy densities of $H$ and $F_3$ diverge at least as\footnote{Evaluating this equation for the case of partially smeared $\overline{\textrm{D}3}$-branes, we recover the result of \cite{Bena:2012vz}, where it was shown that $\e^{2A} \sim \tau^{1/2}$ and $\e^{-\phi} |H|^2 \sim \e^\phi |F_3|^2 \sim \tau^{-1/2}$ near the tip of the conifold and $\tau$ is the radial coordinate transverse to the branes in the conventions of \cite{Bena:2012vz}.}
\begin{equation}
\e^{-\phi} |H|^2 \sim \e^{-2A}, \qquad \e^\phi |F_3|^2 \sim \e^{-2A}.
\end{equation}
Note that, due to its global nature, the argument is independent of most details of the bulk dynamics and does 
therefore not require simplifications such as a partial smearing of the branes or a linearization of the equations 
of motion. Under the assumptions discussed above, it holds for fully localized branes that backreact on the full non-linear equations of motion.
\\

\section{Conclusion}
\label{concl}

We have shown how the 10D equations of motion for classical type II supergravity can be combined to give a surprisingly simple expression for 
the cosmological constant in terms of the classical near-source behavior of the supergravity fields and a contribution from topologically non-trivial background fluxes. The derivation relies on no specific assumptions 
on the compactification manifold, but it holds only for maximally symmetric 
spacetimes of dimension four or more. In simple examples, the flux contribution can be chosen to be zero, and  
 the expression reduces to contributions that have support only  on localized 
sources. This extends the recent work \cite{Burgess:2011rv} to general brane and flux setups.
We checked our result against some well-understood examples of flux compactifications and found agreement with all expectations. We specified the assumptions that are required to apply our result also to de Sitter uplifts from $\overline{\textrm{D}3}$-branes in warped throats and showed that this would then indicate the presence of a singular $H$ and $F_3$ energy density at the $\overline{\textrm{D}3}$-brane similar to what has been reported in recent studies of the same setup \cite{McGuirk:2009xx,Bena:2009xk,Bena:2011hz,Bena:2011wh,Bena:2012bk}. Although our analysis does not clarify the physical meaning of this singularity (see \cite{Bena:2012ek,Blaback:2012nf,Bena:2013hr} for a recent conjecture), it indicates that it is unlikely a mere artifact of approximations such as partial smearing or linearized field equations, which we do not use.

It should be interesting to apply our general result also to other aspects of string compactifications.

\section*{Acknowledgments}
The authors would like to thank Gary Shiu, Yoske Sumitomo and Thomas Van Riet for useful discussions. D. J. would also like to thank the organizers of the workshop ''Brane backreaction, fluxes and meta-stable vacua in string theory`` at Uppsala Universitet for hospitality. This work was supported by the German Research Foundation (DFG) within the Cluster of Excellence "QUEST".

\appendix

\section{Explicit Manipulations of the Equations of Motion}
\label{app_typeII}

Here we present an alternative derivation of our main result \eqref{y}, which only uses the equations of motion. We first consider the Bianchi identity \eqref{bianchix} for the internal RR field strength $F^\inte_{8-p}$ and multiply by $\sigma(C^\ext_{p+1})$,
\begin{eqnarray}  
0 
&=& -\alpha\sigma(C^\ext_{p+1})\w\left\langle \d_{-H}F^\inte + j\right\rangle_{9-p}\notag\\
&=& \d\left[\sigma(C^\ext_{p+1})\w F^\inte_{8-p}\right] + \alpha\langle\sigma(\d_{-H}C^\ext)\rangle_{p+2}\w F^\inte_{8-p} + \alpha\sigma(H\w C^\ext_{p-1})\w F^\inte_{8-p}\notag\\
&& +\alpha\sigma(C^\ext_{p+1})\w H\w F_{6-p}^\inte - \alpha\sigma(C^\ext_{p+1}) \w j_{9-p}\notag\\
&=& \d\left[\sigma(C^\ext_{p+1})\w F^\inte_{8-p}\right] + \alpha\sigma\left\langle F^\ext - \e^B \w F^b \right\rangle_{p+2}\w F^\inte_{8-p} - \alpha\sigma(F_{8-p}^\inte)\w H\w C^\ext_{p-1}\notag\\
&& +\alpha\sigma(F_{6-p}^\inte)\w H\w C^\ext_{p+1}  + \alpha\sigma(j_{9-p})\w C^\ext_{p+1}\notag\\
&=& \d\left[\sigma(C^\ext_{p+1})\w F^\inte_{8-p}\right] - \e^{(p-3)\phi/2}\star_{10}|F_{8-p}^\inte|^2 + \left\langle \e^B \w F^b \right\rangle_{p+2}\w \sigma(F^\inte_{8-p})\notag\\
&& - H\w C^\ext_{p-1}\w\sigma(F^\inte_{8-p}) + H\w C^\ext_{p+1} \w \sigma(F^\inte_{6-p}) - C^\ext_{p+1}\w \sigma(j_{9-p}). \label{bi}
\end{eqnarray}
Here we have introduced the constant $\alpha$ which equals $+1$ for type IIA and $-1$ for type IIB supergravity.
Multiplying the $H$ equation of motion \eqref{h-eoms} by $B$ yields
\begin{eqnarray}  
0 
&=& 2B \w \d\left(\e^{-\phi}\star_{10} H\right) - \alpha\left\langle B\w \sigma(F) \w F\right\rangle_{10}\notag\\
&=& 2B \w \d\left(\e^{-\phi}\star_{10} H\right) - 2\alpha\left\langle B\w \sigma(F^\inte)\w F^\ext\right\rangle_{10}\notag\\
&=& 2\d\left\langle\e^{-\phi}B\w\star_{10} H - B\w\sigma(F^\inte)\w C^\ext\right\rangle_9 - 2(H-H^b)\w \left(\e^{-\phi}\star_{10} H\right)\notag\\
&& + 2\left\langle \d_H(B\w\sigma(F^\inte))\w C^\ext - \alpha B\w\sigma(F^\inte)\w \e^B \w F^b\right\rangle_{10}\notag\\
&=& 2\d\left\langle\e^{-\phi}B\w\star_{10} H - B\w\sigma(F^\inte)\w C^\ext\right\rangle_9 - 2(H-H^b)\w \left(\e^{-\phi}\star_{10} H -  \left\langle\sigma(F^\inte)\w C^\ext\right\rangle_7\right)\notag\\
&&- 2\left\langle   -B\w C^\ext\w\sigma(j) + \e^B \w F^b \w B\w\sigma(F^\inte)\right\rangle_{10}. \label{eom_h}
\end{eqnarray}
Notice in above equation that $F_6^\inte$ never appears since $F^\inte$ is everywhere multiplied by either $B$ or $H$, which must both be purely internal in a maximally symmetric compactification to $d \ge 4$ dimensions. We now take the combination $(1+(p-3)c/2)$ times \eqref{bi} plus $c/2$ times \eqref{eom_h} and sum over $p$. Substituting the definition of $j$ from Section \ref{conventions}, this yields
\begin{eqnarray} 
0 
&=& \sum_{3\le p}\left(1+\f{p-3}{2}c\right)\left\{-\e^{(p-3)\phi/2} \star_{10}|F^\inte_{8-p}|^2  - C^\ext_{p+1}\w\sigma(j_{9-p})\right\}\notag\\
&& + c\left\langle \e^{-\phi}\star_{10}|H|^2 + B\w C^\ext\w\sigma(j) \right\rangle_{10} - \Sigma(c) + \text{total derivatives} \notag\\
&=&\sum_{3\le p} \left(1+\f{p-3}{2}c\right)\left(-\e^{(p-3)\phi/2} \star_{10}|F^\inte_{8-p}|^2 - S_\text{CS}^{(p)}\right)\notag\\
&&  + c\,\e^{-\phi}\star_{10}|H|^2 - \Sigma(c) + \text{total derivatives}, \label{all_bi}
\end{eqnarray}
where $c$ is a free parameter. We also introduced the shorthand
\begin{eqnarray}  
\Sigma (c) &=& -\sum_{2\le p} \left(1+\f{p-3}{2}c\right)F^b_{p+2}\w\langle\e^B\w \sigma(F^\inte)\rangle_{8-p} + \left(1-\f{1}{2}c\right)F^b_4\w \sigma(F^\inte_6) \notag\\
&&- cH^b\w\left( \e^{-\phi}\star H - \langle \sigma( F^\inte)\w C^\ext\rangle_7\right), \label{sigma2}
\end{eqnarray}
where we have combined all terms that depend on background fluxes to simplify our notation.

The trace of the external components of the (trace-reversed) Einstein equation reads
\begin{equation}
\frac{4}{d} R_d =  - \frac{1}{2} \e^{-\phi} |H|^2 + \sum_{3\le p} \f{p-7}{4}\left(\e^{(p-3)\phi/2} |F_{8-p}^\inte|^2  \pm \mu_p \e^{(p-3)\phi/4} \delta (\Sigma)\right) + \f{5}{4} \e^{\phi/2}|F_4^\ext|^2\text, \label{einstein-app}
\end{equation}
where the upper sign is for D-branes and the lower sign for O-planes and we have used $|F_5^\ext|^2 = -|F_5^\inte|^2$ to rewrite the spacetime-filling part of $|F_5|^2$. Note that spacetime-filling $F_4$ flux can only be present for $d=4$ in type IIA supergravity, while $F_5$ flux can be present for $d=4$ or $d=5$ in type IIB supergravity.

The dilaton equation \eqref{dilaton} yields
\begin{equation}
0 =  -\nabla^2 \phi - \frac{1}{2} \e^{-\phi} |H|^2 + \sum_{3\le p} \frac{p-3}{4}\left( \e^{(p-3)\phi/2} |F^\inte_{8-p}|^2 \pm \mu_p \e^{(p-3)\phi/4} \delta(\Sigma)\right) + \f{1}{4}\e^{\phi/2}|F_4^\ext|^2. \label{dilaton-app}
\end{equation}
Combining \eqref{einstein-app} and \eqref{dilaton-app}, we find
\begin{align}
\frac{4}{d} R_d &=  c\, \e^{-\phi} |H|^2 + \sum_{3\le p} \left({1+\frac{p-3}{2}c}\right)\left(-\e^{(p-3)\phi/2} |F^\inte_{8-p}|^2  \mp \mu_p \e^{(p-3)\phi/4} \delta(\Sigma)\right) \notag \\ &\quad + \left({1-\f{c}{2}}\right) \e^{\phi/2}|F_4^\ext|^2 +\text{total derivatives}. \label{es_dil}
\end{align}
Finally, we can combine \eqref{es_dil} with \eqref{all_bi} to get
\begin{align}
\frac{4}{d} \star_{10} R_d &=  \sum_{3\le p}\left({1+\frac{p-3}{2}c}\right)\left(\mp \star_{10}\mu_p \, \e^{(p-3)\phi/4} \delta(\Sigma) + S_\text{CS}^{(p)} \right) +\mathcal{F}(c) \notag \\ &\quad +\text{total derivatives}, \label{es_dil_IIa}
\end{align}
where we defined
\begin{equation}
\mathcal{F}(c) = \Sigma(c) - \left(1 - \f{c}{2}\right) F_4^b \w \sigma( F^\inte_6) \label{sigma3}
\end{equation}
and used $\e^{\phi/2} \star_{10} F_4^\ext = - \sigma( F^\inte_6)$, which follows from the duality relations \eqref{rrduality}.
Integrating over ten-dimensional space and using \eqref{ricci-lambda}, we get rid of all total derivative terms and find
\begin{equation}
\frac{8v\mathcal{V}}{d-2} \Lambda = \sum_p\left({1+\frac{p-3}{2}c}\right) \left[S^{(p)}_\text{DBI} + S_\text{CS}^{(p)}\right] + \int\mathcal{F}(c),
\end{equation}
with the volume factors $v$ and $\mathcal{V}$ defined as in \eqref{volumes}.
\\

\bibliographystyle{utphys}
\bibliography{groups}

\end{document}